\documentclass[aps,
prb,
reprint,
preprintnumbers,
superscriptaddress,
amsfonts,
amssymb,
amsmath
]{revtex4-2}
\usepackage{bm,latexsym,mathrsfs,enumerate}
\usepackage{upgreek}
\usepackage[table,x11names]{xcolor}
\usepackage[breaklinks=true,
unicode=true,
urlcolor = blue,
colorlinks = true,
citecolor = blue,
linkcolor = blue
]{hyperref}
\usepackage{graphicx}
\graphicspath{{./figs/}}
\usepackage{savesym}
\usepackage{lmodern}
\usepackage{mathtools}
\savesymbol{tensor}
\usepackage{mismath}
\usepackage{pifont}
\usepackage{multirow}
\renewcommand{\vec}[1]{\bm{#1}}
\DeclareMathOperator{\sign}{sign}
\DeclareMathAlphabet{\mathpzc}{OT1}{pzc}{m}{it}
\begin{document}

\title{Chiral breakdown engineered by mesoscale Dzyaloshinskii-Moriya interaction\\
in biaxial magnetic nanotubes}
	
\author{Kostiantyn V. Yershov}
\email{k.yershov@ifw-dresden.de}
\affiliation{Institute for Theoretical Solid State Physics, Leibniz Institute for Solid State and Materials Research Dresden, D-01069 Dresden, Germany}
\affiliation{Bogolyubov Institute for Theoretical Physics of the National Academy of Sciences of Ukraine, 03143 Kyiv, Ukraine}

\author{Svitlana Kondovych}
\email{s.kondovych@ifw-dresden.de}
\affiliation{Institute for Theoretical Solid State Physics, Leibniz Institute for Solid State and Materials Research Dresden, D-01069 Dresden, Germany}

\author{Denis D. Sheka}
\email{sheka@knu.ua}
\affiliation{Taras Shevchenko National University of Kyiv, 01601 Kyiv, Ukraine}

\date{February 12, 2025}
	
\begin{abstract}

Curvilinear geometries in magnetic nanostructures provide a unique platform for exploring the interplay of symmetry, topology, and curvature in magnetization dynamics. In this work, we analytically study the static and dynamic properties of domain walls in biaxial magnetic nanotubes with intrinsic Dzyaloshinskii–Moriya interaction of different symmetries. We show that geometry-driven local and nonlocal interactions govern domain profiles and dynamics, enabling precise control over the wall propagation and Walker breakdown field. Furthermore, the combination of bulk-type Dzyaloshinskii–Moriya interaction and curvature leads to chirality symmetry breaking and chiral breakdown in domain wall motion. These findings offer a framework for tailoring domain wall textures in cylindrical nanotubes, unlocking new functionalities for advanced applications in curvilinear magnonics and data storage technologies.
\end{abstract}

\maketitle

\section{Introduction}
\label{sec:intro}

Domain walls (DWs) in ferromagnetic nanostructures represent a cornerstone of modern magnetism, with significant implications for spintronic and data storage technologies~\cite{Allwood02,Allwood05,Parkin08,Luo20,Yang21,Venkat23}. Ferromagnetic thin films~\cite{Thiaville12,Boulle13a,Emori13,Ryu13,Brataas13,Kravchuk14,Yershov20b,Pylypovskyi20b}, nanowires~\cite{Thiaville05,Yan10,Yershov15b,Yershov16,Pylypovskyi16,Fernandez-Roldan22}, and nanotubes~\cite{Yan11a,Landeros10,Yan12,Otalora12a,Otalora12,Otalora13,Goussev16,Depassier19,MancillaAlmonacid20,Josten21,Landeros22} exhibit a wide variety of DW configurations, including Bloch and N{\'e}el walls, as well as more exotic hybrid textures driven by the Dzyaloshinskii--Moriya interaction (DMI)~\cite{Dzyaloshinsky58,Moriya60}. The specific properties of these DWs -- such as their magnetization profile, stability, mobility, and chirality -- are determined by the underlying geometry and interactions within the material, allowing for tailored functionalities in advanced applications.

The stabilization of DWs depends on a complex interplay of forces, with key contributions from exchange interactions, magnetic anisotropy, DMI, and magnetostatic (dipole-dipole) interactions. In particular, local intrinsic DMI induces non-trivial chiral spin textures, such as skyrmions and chiral DWs, while magnetostatics shapes the overall energy landscape that alters the texture profile and influences its dynamics~\cite{Woo16,Lemesh17,Buettner18,BernandMantel18,Schoepf23}. In confined systems, nonlocal dipole-dipole interactions can  give rise to chiral structures even in the absence of DMI~\cite{Lemesh17,Buettner18}. However, in thin-film geometries, magnetostatics merely contributes to the local magnetic anisotropy energy~\cite{Carbou01,Kohn05}, thereby primarily affecting the DW width. 

In systems with curvilinear geometries, curvature and topology offer numerous intriguing effects which enrich the magnetic texture scenery~\cite{Sheka22b,Sheka23,Sheka20a}.  The curvature influences both static and dynamics of magnetization textures. The most prominent manifestation of curvature-induced effects are topological patterning and geometric magnetochiral effects~\cite{Sheka21}. A paradigmatic example of curvature-induced effects is offered by magnetic nanotubes~\cite{Landeros10,Yan11a,Yan12,Otalora12,Otalora12a,Otalora13,Goussev16,Depassier19,MancillaAlmonacid20,Josten21,Landeros22}, they have become experimentally accessible through advanced fabrication techniques~\cite{Proenca20,Streubel21,Landeros22} and offer an important platform for next-generation technologies, including magnetic sensors, memory elements, and spintronic devices~\cite{Makarov22a}. Magnetic nanotubes offer enhanced stabilization and control over DWs, providing new degrees of freedom through geometry-induced effects on DMI, anisotropy, and magnetostatic contributions. For instance, pattern-induced chiral symmetry breaking can occur in nanotubes due to the dipole–dipole interaction: two energetically equivalent vortex DWs with opposite chiralities exhibiting different dynamic properties~\cite{Landeros10,Otalora12a,Otalora13}, leading to the suppression of Walker breakdown~\cite{Schryer74} and the Cherenkov-like radiation of magnons for fast DWs~\cite{Yan11a,Yan13}. The nonlocal dipole-dipole interaction in the vortex state tube introduces effects in magnetization dynamics not present in flat systems of the same material, such as asymmetric spin-wave propagation~\cite{Otalora16,Brevis24}. The joint action of intrinsic DMI and geometry-governed exchange-driven DMI give rise to mesoscale DMI providing different opportunities for geometrical manipulations of material responses~\cite{Volkov18}. 
In tubular geometry, mesoscale DMI becomes a source of a new type of inclined DWs \cite{Yershov20, Liu20b} and high velocities of vortex DWs~\cite{Goussev16,MancillaAlmonacid20} in magnetic nanotubes subjected to magnetic fields and spin-polarized currents.

In this work, we demonstrate the profound influence of curvature on the static and dynamic properties of DWs in biaxial magnetic nanotubes with intrinsic DMI of different symmetries. We establish that the interplay between bulk-type DMI and curvature results in chirality symmetry breaking, driven by a preferred combination of DW topological charge and helicity. In a vortex magnetic field, this causes a chiral breakdown in DW propagation, whereas in an axial magnetic field it defines the direction of the DW motion.

We further show that for nanotubes with interfacial-type DMI, the DW profile and dynamics are governed by the geometry-induced competition between local and nonlocal effects. While this symmetry configuration restrains chiral effects, it allows for tailored control over the Walker limit in an external magnetic field.

\section{Model of a magnetic nanotube}
\label{sec:model}

\begin{figure}[t]
\includegraphics[width=\columnwidth]{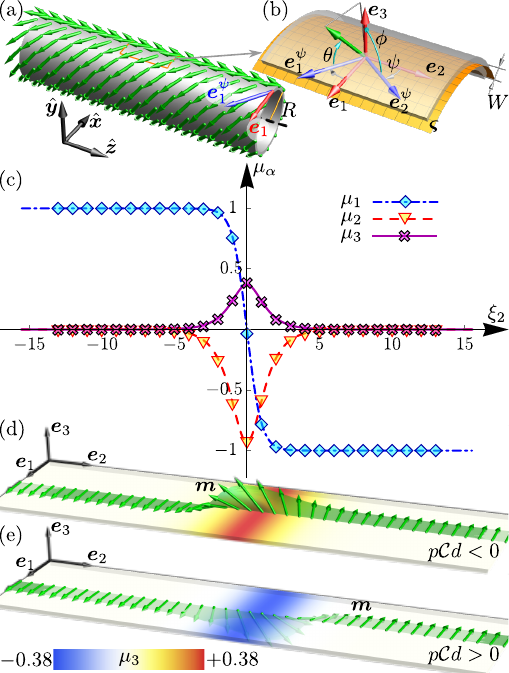}
\caption{\textbf{Static domain walls in a cylindrical shell with bulk-type DMI.} (a),(b) Schematic representation of a magnetic texture on a cylindrical tubular shell of mean radius $R$ and thickness $W$. Magnetization $\vec{m}$ (green arrows) forms a tilted vortex state at equilibrium (a) and is parametrized by angles $\left\{\theta,\phi\right\}$ in $\psi$-rotated reference frame (b). (c) Numerical simulation results (symbols) for magnetization components $\mu_\alpha$ in the rotated reference frame~\eqref{eq:b_magn}. Lines correspond to the analytical ansatz~\eqref{eq:b_dw_solution} with equilibrium DW profile parameters~\eqref{eq:b_equilibrium}. Panel (d) illustrates the reconstructed magnetization distribution from (c) in an unrolled cylindrical shell with $p\,\mathcal{C}d<0$. Color scheme corresponds to the normal to the surface magnetization component $\mu_3$. Green arrows show the DW magnetic texture, $\vec{m}(\xi_2)$, in which $\mu_3$ is amplified (2 times) for visibility due to its small magnitude. Parameters used in numerical simulations are $p=1$, $d = 0.25$, $\varepsilon = 0.5$, and $\varkappa = 0.2$, resulting in the tilt of $\vec{m}$ by a small angle $\psi \approx 0.025$. (e) DW with the opposite chirality, $p\,\mathcal{C}d > 0$.
\label{fig:bDW}
}
\end{figure}

We consider a ferromagnetic cylindrical tubular shell of constant thickness $W$ and mean radius $R$, described by the curved surface $\vec{\varsigma}(x_1,x_2) =\vec{\gamma}(x_1)+x_2\hat{\vec{z}}$ with $\vec{\gamma}=R\cos\left(x_1/R\right)\hat{\vec{x}}+R\sin\left(x_1/R\right)\hat{\vec{y}}$, see Fig.~\ref{fig:bDW}(a),(b). Here, $x_1$ and $x_2$ are curvilinear coordinates on the surface $\vec{\varsigma}$. The introduced parameterization $\vec{\varsigma}(x_1,x_2)$ induces Eucledian metric $g_{\alpha\beta}=\delta_{\alpha\beta}$ and the  tangential basis $\vec{e}_\alpha = \partial_\alpha\vec{\varsigma}$, where $\vec{e}_1 = \partial_1 \vec{\gamma}$ is the unit vector tangential to $\vec{\gamma}$ and $\vec{e}_2 = \hat{\vec{z}}$. The basis $\{\vec{e}_1,\vec{e}_2\}$ is orthonormalized, thus the unit surface normal is introduced as $\vec{e}_3=\vec{e}_1\times\vec{e}_2$. This is a particular case of the cylinder surface parametrization~\cite{Yershov20}, which corresponds to the conventional cylindrical coordinate system.

Curvilinear properties of the surface $\vec{\varsigma}$ are completely determined by the only parameter: curvature $\kappa = 1/R$ of the curve $\vec{\gamma}$.  We assume that the thickness $W$ of the film is small enough to ensure that the magnetization $\vec{M}$ is uniform along normal direction $\vec{e}_3$ (coordinate $x_3$). The energy of such a magnetic tube we model with the following functional,
\begin{equation}\label{eq:model}
	\mathcal{E}[\vec{m}]=\frac{E}{E_0}=\iint\left[\mathscr{E}_\textsc{x}+d\mathscr{E}_\textsc{d}+\mathscr{E}_\textsc{a}\right]\textrm{d}\xi_1\mathrm{d}\xi_2 + \frac{\mathcal{E}_\textsc{ms}}{w Q},
\end{equation}
where $\vec{m}=\vec{M}/M_\text{s}$, $M_\text{s}$ is the saturation magnetization, $E_0 = K_\text{a} \ell^3 w$ with $K_\text{a}>0$ being the easy-axial (along $\vec{e}_1$) anisotropy constant arising from sample growth \cite{Josten21,Zimmermann18}, $\ell = \sqrt{A/K_\text{a}}$ is the characteristic length scale of the system (magnetic length), with $A$ being exchange stiffness, $\xi_\alpha = x_\alpha /\ell$ are dimensionless curvilinear coordinates, $w = W /\ell$ is the dimensionless film thickness, $Q=2K_\text{a} /(\mu_0 M_\text{s}^2)$ is the anisotropy quality factor of a material, and $\mu_0$ is the magnetic permeability of vacuum. The first integral part on the right-hand side of~\eqref{eq:model} comprises the local energy terms. The first term in square brackets is energy density of the non-uniform exchange, $\mathscr{E}_\textsc{x}=-\vec{m}\cdot\vec{\nabla}^2\vec{m}$. The second term corresponds to the DMI energy density $\mathscr{E}_\textsc{d}$, with $d=D/\sqrt{AK_\text{a}}$ being the dimensionless DMI constant. We consider two types of DMI: (i) $\mathscr{E}_\textsc{d}^b = \vec{m}\cdot\left[\vec{\nabla}\times\vec{m}\right]$ is applicable for systems with $T$ and $O$ symmetries~\cite{Cortes-Ortuno13}. In the following this is called bulk-type DMI. (ii)~$\mathscr{E}_\textsc{d}^i=- \vec{m}\cdot \left[\left(\vec{e}_3\times\vec{\nabla}\right) \times \vec{m}\right]$ is valid for ultrathin films~\cite{Bogdanov01,Thiaville12}, bilayers~\cite{Yang15}, or materials belonging to $C_{nv}$ crystal classes. In the following we call this DMI of interfacial type. Here and below indices $b$ and $i$ correspond to the bulk and interfacial DMI types, respectively. The third term in square brackets corresponds to the biaxial anisotropy $\mathscr{E}_\textsc{a}=1 - \left(\vec{m}\cdot\vec{e}_1\right)^2 + \epsilon\, \left(\vec{m}\cdot\vec{e}_3\right)^2$ with anisotropy ratio $\epsilon = K_\text{s}/K_\text{a}>0$, where $K_\text{s}$ is the easy-surface anisotropy constant. The case with $\epsilon > 1$ corresponds to the nanotubes, experimentally studied by Josten~\textit{et al.}~\cite{Josten21}.  The last energy term in \eqref{eq:model} describes the nonlocal magnetostatic energy due to dipole-dipole interactions, $\mathcal{E}_\textsc{ms} = \int \mathrm{d}\vec{\xi} \int \mathrm{d}\vec{\xi'} \left(\vec{m}(\vec{\xi})\cdot \vec{\nabla} \right)\left(\vec{m}(\vec{\xi'})\cdot \vec{\nabla'} \right) {\lvert \vec{\xi}-\vec{\xi'} \rvert}^{-1}$.

In the following sections, we separately consider the two types of DMI, deriving the DW magnetization profiles that minimize energy functional~\eqref{eq:model} for each case, and then analyze the DW dynamics in an external magnetic field.

\section{Domain wall on a nanotube \\ with bulk-type DMI}
\label{sec:bDMI}

In the absence of DMI, the equilibrium magnetization texture forms a vortex state with $\vec{m}_{\text{v}}(\xi_1) = \vec{e}_1$~\cite{Zimmermann18}. Stabilized mainly due to anisotropy $K_a$, the vortex state is favorable for $\varkappa <2\sqrt{1+\varepsilon}/\pi$, see Appendix~\ref{app:param}. The symmetry of bulk-type DMI, $\mathscr{E}_\textsc{d}^b = \vec{m}\cdot\left[\vec{\nabla}\times\vec{m}\right]$, favors the emergence of the magnetization component $m_3$, normal to the surface of the nanotube. The DMI also inclines the magnetization, leading to the \emph{tilted vortex state}, see Fig.~\ref{fig:bDW}(a). The direction of this deviation, determined by a tilt angle $\psi \approx -d\varkappa / 2$, depends on the sign of DMI constant $d$, see Appendix~\ref{app:bDMI_psi_frame} for details. Note that the tilt is proportional to the dimensionless curvature $\varkappa = \ell/R$.

To parameterize the unit vector of magnetization, we first rotate the reference frame in the local rectifying surface by the angle $\psi$, as shown in Fig.~\ref{fig:bDW}(b). The magnetization in the rotated $\psi$-frame~$\{\vec{e}_1^\psi,\vec{e}_2^\psi,\vec{e}_3^\psi\equiv\vec{e}_3\}$ reads
\begin{equation}\label{eq:b_magn}
	\vec{m} = \mu_\alpha\vec{e}_\alpha^\psi = \cos\theta\vec{e}_1^\psi+\sin\theta\left(\cos\phi\vec{e}_2^\psi+\sin\phi\vec{e}_3\right),
\end{equation}
where angular variables $\theta$ and $\phi$ depend on the spatial and temporal coordinates. Using this reference frame, we diagonalize the effective anisotropy energy density of the cylinder (Appendix~\ref{app:bDMI_psi_frame}). The total local energy density in the rotated frame of reference reads ($\{\alpha,\beta\}=\{1,2\}$):
\begin{equation}\label{eq:b_energy_rotated}
	\begin{split}
		\mathscr{E}^b = \partial_\alpha\mu_\beta\partial_\alpha\mu_\beta&-\mathcal{K}^b_1 \mu_1^2 + \mathcal{K}^b_3\mu_3^2\\
		&+ \mathcal{D}_{\beta 3}^{b(\alpha)}\left(\mu_\beta\partial_\alpha\mu_3-\mu_3\partial_\alpha\mu_\beta\right).
	\end{split}
\end{equation}
Coefficient $\mathcal{K}^b_1$ characterizes the strength of the effective easy-axis anisotropy, while $\mathcal{K}^b_3$ gives the strength of the effective easy-surface anisotropy. The parameters $\mathcal{D}_{\beta 3}^{b(\alpha)}$ can be interpreted as mesoscale DMI coefficients~\cite{Volkov18}. In the case of small curvature and DMI strength, the effective anisotropy and DMI constants can be attributed to the geometrical parameters of the system:
\begin{equation}\label{eq:b_const_rotated}
	\begin{split}
		&\mathcal{K}^b_1\approx 1- \varkappa^2\left(1-d^2/2\right)\!,\quad \mathcal{K}^b_3\approx {\varepsilon} + \varkappa^2\left(1-d^2/4\right)\!,\!\\
		&\mathcal{D}_{1 3}^{b(1)} \approx -2\varkappa+\varkappa d^2/2,\quad \mathcal{D}_{1 3}^{b(2)} \approx d,\\
		&\mathcal{D}_{2 3}^{b(1)} \approx -d - d\varkappa^2,\quad \mathcal{D}_{2 3}^{b(2)} \approx \varkappa d^2/2.\\
	\end{split}
\end{equation}

The nonlocal magnetostatic energy term in~\eqref{eq:model}, $\mathcal{E}_\textsc{ms}$, comprises the interactions between the surface, $\sigma^{\pm}=\pm m_3$, tangential, $\rho=-\partial_1 m_1-\partial_2 m_2$, and geometrical, $g =-m_3/x_3$, magnetostatic charges~\cite{Sheka20a}.  In the considered tubular geometry, both anisotropy and magnetostatics favor the uniform tangential magnetization (competing with the DMI contribution), and thus it is only the DW region that is charged. The $\psi$-rotation of the reference frame does not affect $\sigma$ nor $g$ since $\mu_3=m_3$, and the tangential charge is rotation-invariant, $\rho=-\partial_1 \mu_1-\partial_2 \mu_2$. 

Here, we consider the thin-shell limit case, $w\varkappa \ll 1$, in which the leading contribution to magnetostatic energy comes from the surface charges and is local and linear in $w$. All other interactions are nonlocal and scale as higher order on the shell thickness $w$~\cite{Carbou01,Kohn05,Fratta20a}. Moreover, the magnetic symmetry of terms corresponding to interactions $\mathcal{E}_\textsc{ms}^{\rho-\rho}$, $\mathcal{E}_\textsc{ms}^{g-g}$, and $\mathcal{E}_\textsc{ms}^{\sigma-g}$ restricts symmetry breaking and cannot lead to chiral effects, resulting just in a small correction of $\lito(w)$ to the total energy. Based on these considerations, we include the local part of magnetostatic energy to the energy density~\eqref{eq:b_energy_rotated} as $\mathcal{E}_\textsc{ms}^{\sigma-\sigma}=w\iint \mu_3^2 \textrm{d}\xi_1\mathrm{d}\xi_2$ (see~Appendix~\ref{app:bDW_ms}) by explicitly shifting the anisotropy coefficient  in~\eqref{eq:b_const_rotated}: $\varepsilon=\epsilon+Q^{-1}$. Further, we derive the possible static magnetization configurations by variation of the total energy functional with density~\eqref{eq:b_energy_rotated}, and discuss the remaining magnetostatic terms $\mathcal{E}_\textsc{ms}^{\sigma-\rho}$ and $\mathcal{E}_\textsc{ms}^{g-\rho}$ based on the symmetry of the obtained variational ansatz.   

\subsection{Static domain wall solution}\label{sec:bDMI_static}

The ground state magnetization textures arising from energy functional~\eqref{eq:model} with bulk-type DMI form two tilted vortices with $\cos\theta_\text{tv} = C =\pm1$, which correspond to the two possible directions of the vortex circulation: clockwise with $C=1$ (see Fig.~\ref{fig:bDW}(a)) and counterclockwise with $C=-1$. The structure of a DW between two tilted vortices with opposite circulations can be described analytically with the following ansatz~(see Appendix~\ref{app:bDW_sol}):
\begin{equation} \label{eq:b_dw_solution}
\cos\theta_\textsc{dw}^b = -p \tanh\frac{\xi_2}{\Delta},\quad \phi_\textsc{dw}^b = \varPhi.
\end{equation}
Here, $p = \pm 1$ is a topological charge of the DW, and $\Delta = 1/\sqrt{\mathcal{K}^b_1}$ is a DW width. In the absence of DMI, ansatz~\eqref{eq:b_dw_solution} with $\cos \Phi_0=\mathcal{C}=\pm1$ becomes an exact solution, see Appendix~\ref{app:bDW_sol}. In the following we proceed with the investigation of the finite curvature effects on the magnetization distribution in DWs in nanotubes. We will apply a variational approach by using~\eqref{eq:b_dw_solution} as a DW ansatz with the DW width $\Delta$ and initial phase $\varPhi$ being the variational parameters. By substituting~\eqref{eq:b_dw_solution} into the energy density functional~\eqref{eq:b_energy_rotated} with account of surface magnetostatic energy term and integrating over the $\xi_1$ and $\xi_2$,  we obtain
\begin{equation}\label{eq:b_dw_energy}
	\frac{\mathcal{E}_\textsc{dw}^b}{2\pi/\varkappa} = \frac{2}{\Delta}+2\Delta\left[\mathcal{K}^b_1\!+\mathcal{K}^b_3\sin^2\varPhi\right]+p\,\pi\mathcal{D}_{13}^{b(2)}\sin\varPhi.
\end{equation}
The first two terms on the right-hand side in~\eqref{eq:b_dw_energy} determine the competition of the non-uniform exchange and anisotropy contributions, while the third term originates from the mesoscale DMI and demonstrates the coupling between the DMI strength $\mathcal{D}_{13}^{b(2)}$, DW topological charge $p$, and helicity $\mathcal{C}$. By introducing the DW phase $\varPhi = \varPhi_0 + \tilde{\varPhi}$ one can write the last term as $p\,\mathcal{C}\pi\mathcal{D}_{13}^{b(2)}\sin\tilde{\varPhi}$.  This term depends on the chirality of a DW, defined by the sign of the product $\left(p\,\mathcal{C}d\right)$.

\begin{figure*}[t]
\includegraphics[width=\textwidth]{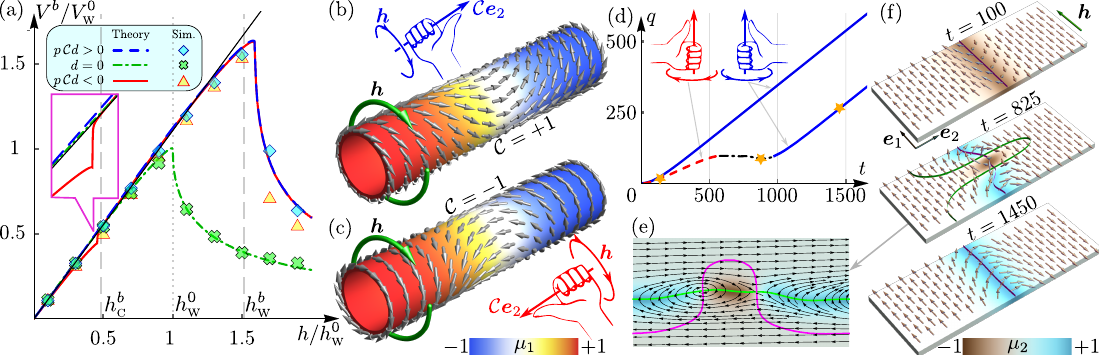}
\caption{\textbf{DW motion in the applied vortex magnetic field in nanotubes with bulk-type DMI.} (a) DW velocity $V^b$ as a function of applied vortex magnetic field $\vec{h} = h \vec{e}_1$. Velocity is measured with respect to DMI-free Walker velocity $V_\textsc{w}$, and the value of the magnetic field is given in units of DMI-free Walker field $h_\textsc{w}^0$~\eqref{eq:b_walker_field} (dotted vertical light gray line). Solid black line is the DW velocity in a traveling-wave regime~\eqref{eq:bDW-vel-V}. Colored lines correspond to the numerical solution of the equations of motion~\eqref{eq:b_motion_vortex}. Dashed vertical light gray lines mark the analytically predicted values for the chiral breakdown field $h_\textsc{c}^b$ and Walker field $h_\textsc{w}^b$. The inset shows the enlarged area of the plot where the chiral breakdown occurs. Symbols correspond to the results of numerical simulations (Appendix~\ref{app:sim}). Panels (b) and (c) illustrate nanotubes with DWs of opposite helicities, $\mathcal{C}=\pm 1$. The hand icons visualize the interplay between the DW helicity and the magnetic field: the thumbs point in the direction of $\mathcal{C}\,\vec{e}_2$, while the fingers curl along the orientation of the vortex field, $\vec{h}$. (d)~Time evolution of the DW position, $q$, for DWs of different helicities in the vortex magnetic field $h/h_\textsc{w}^0 = 0.7$, extracted from numerical simulations. Solid blue lines correspond to the DW with favorable helicity, $\mathcal{C}=+1$ (b), dashed red line to the DW with $\mathcal{C}=-1$ (c), and dashdotted black line indicates the transient stage. (e) Formation of a vortex-antivortex pair during the transient stage, shown by magnetization streamlines. (f) Snapshots of the magnetization texture of the unrolled nanotube before ($t=100$), during ($t=825$), and after ($t=1450$) the DW helicity switching. Three time points are shown in panel (d) with orange stars. The middle snapshot captures the vortex-antivortex pair mediated breakdown shown in panel (e). Green and magenta lines in panels (f) and (e) correspond to isosurfaces with $\mu_1=0$ and $\mu_2=0$, respectively. Parameters used in numerical simulations are $p=\pm 1$, $d=0.25$, $\varkappa = 0.2$, $\eta = 0.01$, and $\varepsilon = 0.5$. Color schemes in (b) and (e,f) correspond to the $\mu_1$ and $\mu_2$ magnetization components, respectively. 
\label{fig:vel_bDMI}}
\end{figure*}

With variational ansatz~\eqref{eq:b_dw_solution} we can estimate the remaining nonlocal contributions to magnetostatic energy $\mathcal{E}_\textsc{ms}^{\sigma-\rho}$ and $\mathcal{E}_\textsc{ms}^{g-\rho}$, which favor the coupling between the normal magnetization component $\mu_3$ and the gradient of the tangential component $\partial_2 \mu_2$, potentially allowing the chiral effects, such as dynamic chiral symmetry breaking effects observed in~\cite{Landeros10,Otalora12a,Yan12} and static nonlocal chiral symmetry breaking effect observed in \cite{Volkov23}. However, in the DW configuration on a nanotube with bulk-type DMI (Fig.~\ref{fig:bDW}), the components $\mu_2$ and $\mu_3$ have the same parity with respect to $\xi_2\rightarrow -\xi_2$, so energy terms $\mathcal{E}_\textsc{ms}^{\sigma-\rho}$ and $\mathcal{E}_\textsc{ms}^{g-\rho}$ are the same for the DWs of different chiralities shown in Fig.~\ref{fig:bDW}(d),(e) and Fig.~\ref{fig:vel_bDMI}(b),(c). Since no nonlocal chiral effects are expected, and the order of these energy terms is $\lito(w)$~\cite{Carbou01,Kohn05a,Kohn05,Fratta20a}, we consider for this case only the local surface magnetostatic contribution $\mathcal{E}_\textsc{ms}^{\sigma-\sigma}$ as described above.

The minimization of~\eqref{eq:b_dw_energy} with respect to the DW width and phase results in the following equilibrium values
\begin{equation}\label{eq:b_equilibrium}
	\begin{split}
		&\Delta^b_0 = \sqrt{\frac{1}{\mathcal{K}^b_1}-\frac{1}{\mathcal{K}^b_1\mathcal{K}^b_3}\left[\frac{\mathcal{D}_{13}^{b(2)}}{d_0}\right]^2},\\
		&\sin\varPhi^b_0 = - p\,\frac{\mathcal{D}_{13}^{b(2)}}{d_0}\frac{\sqrt{\mathcal{K}_1^b/\mathcal{K}_3^b}}{\sqrt{\mathcal{K}_3^b-\left[\mathcal{D}_{13}^{b(2)}/d_0\right]^2}},
	\end{split}	
\end{equation}
where $d_0 = 4/\pi$ is the critical value of DMI which separates homogeneous and inhomogeneous magnetization textures in planar easy-axial systems~\cite{Rohart13}. The case of small curvature and DMI strength corresponds to $\Delta_0^b\approx 1+\left[\varkappa^2-\frac{\left(d/d_0\right)^2}{\varepsilon}\right]/2$ and $\varPhi^b_0 \approx \varPhi_0 - p\,\mathcal{C} \frac{d}{d_0\varepsilon}$, valid for $d/d_0<\varepsilon$. The comparison of these predictions with numerical simulations is in good agreement, see Fig.~\ref{fig:bDW}(c).

\subsection{Domain wall dynamics \\ in an external magnetic field}\label{sec:bDMI_dynamics}

The interaction of a magnetic texture with an external magnetic field is described by the Zeeman energy density $\mathscr{E}_\textsc{z}= - 2\vec{h}\cdot\vec{m}= - 2\vec{h}^\psi\cdot\vec{\mu}$, where $\vec{h} = \vec{H} / H_\textsc{a}$ is normalized external magnetic field with $H_\textsc{a} = 2 K_a/M_s$ being anisotropy field.

To derive effective equations of the DW motion we use collective variable approach based on a  $q-\Phi$ model~\cite{Slonczewski72}
\begin{equation}\label{eq:b_q-phi}
	\cos\theta^b_\textsc{dw}\! =\! -p \tanh\frac{\xi_2-q(t)}{\Delta},\!\quad \phi^b_\textsc{dw} \!=\varPhi(t),
\end{equation}
where $q$ and $\varPhi$ are time-dependent collective variables. Here and below we consider dimensionless time measured in units of $\gamma_0 K_a/M_s$, where $\gamma_0$ is the gyromagnetic ratio.

\subsubsection{The case of a vortex magnetic field}\label{sec:bDMI_vField}
First, we consider the dynamics of DWs in a nanotube in a vortex magnetic field. The magnetic field is oriented in the tangential direction to the nanotube's surface, $\vec{h} = h \vec{e}_1$, which can be realized as {\O}rsted field induced by a current flowing along the tube symmetry axis. In the rotated reference frame, $\vec{h}^\psi = h \left(\cos\psi \vec{e}_1^\psi - \sin\psi \vec{e}_2^\psi\right)$.  By substitution of ansatz~\eqref{eq:b_dw_solution} into the Zeeman energy and integration over $\xi_1$ and $\xi_2$ we obtain
\begin{equation}\label{eq:b_zeeman_vortex}
	\frac{\mathcal{E}_\textsc{z}^b}{2\pi/\varkappa} = -2h\left(2 p q\cos\psi - \pi \Delta \cos\varPhi\sin\psi \right).
\end{equation}
Here, the first term ($\propto h\cos\psi$) pushes the DW along the tube axis, while the second term ($\propto h\sin\psi$) results in the deformation of the DW profile. However, in our case, we have $|\sin\psi|\approx|d\varkappa/2|\ll 1$, and as a result we can neglect the dynamical changes of the DW width.

In terms of the collective variables, the equations of motion take the form (for details see Appendix~\ref{app:bDW_exat})
\begin{equation}\label{eq:b_motion_vortex}
	\begin{split}
		\frac{\eta}{\Delta}\dot{q} + p\, \dot{\varPhi} = 2ph\cos\psi,&\\
		p\dot{q} - \eta\Delta\dot{\varPhi} = 2p\frac{\mathcal{D}_{13}^{b(2)}}{d_0}&\cos\varPhi+\mathcal{K}_3^b\Delta\sin2\varPhi\\
		&-\pi h\Delta\sin\varPhi\sin\psi.
	\end{split}
\end{equation}
Here, overdot denotes the time derivative with respect to dimensionless time, and $\eta$ is the Gilbert damping parameter. The DW width is assumed to be a slave variable~\cite{Hillebrands06}, i.e., $\Delta(t)=\Delta\left[\varPhi(t)\right] = 1/\sqrt{\mathcal{K}_1^b+\mathcal{K}_3^b\sin^2\varPhi^b}$. The behavior of the DW width is discussed in detail in Appendix~\ref{app:bDW_exat}. 

Set of equations \eqref{eq:b_motion_vortex} has a traveling-wave regime with $\dot{q}=V^b=\text{const}$ and  $\dot{\varPhi^b} = 0$. The corresponding DW velocity and phase are
\begin{subequations}\label{eq:b_trav_vortex}
	\begin{align}\label{eq:bDW-vel-V}
		V^b &\!= 2p\,\frac{h\,\Delta_0^b}{\eta}\cos\psi\!\approx 2\frac{p\,h}{\eta}\left[1+\frac{1}{2}\!\left(\varkappa^2-\frac{1}{\varepsilon}\frac{d^2}{d_0^2}\right)\right],\\
		\label{eq:bDW-phase-V}
		\varPhi^b &\approx \varPhi_0^b +\frac{h}{\eta\varepsilon}\left[1-\frac{\varkappa^2}{\varepsilon}+\frac{d}{d_0\varepsilon}\left(\frac{d}{d_0\varepsilon}-\mathcal{C}h\varkappa\right)\right].
	\end{align}
\end{subequations}

The resulting DW velocity (measured with respect to DMI-free Walker velocity $V_\textsc{w}=\left(\varepsilon+\varkappa^2\right)\sqrt{2}/\sqrt{2+\varepsilon-\varkappa^2})$ as a function of the applied field is presented in~Fig.~\ref{fig:vel_bDMI}(a). The traveling-wave solution exists for the fields $h<h_\textsc{w}^b$, where
\begin{equation}\label{eq:b_walker_field}
	h_c^b \approx h_\textsc{w}^0 + p\,\mathcal{C}\frac{\eta}{\sqrt{2}} \frac{d}{d_0},\qquad h_\textsc{w}^0 = \frac{\eta}{2}\left(\varepsilon + \varkappa^2\right).
\end{equation}
Here, $h_\textsc{w}^0$ is the Walker field for a DMI-free biaxial ferromagnetic nanotube. For a fixed DW topological charge and a sign of DMI strength~(\textit{e.g.} for $\sign(pd)=+1$) we can introduce two critical fields, namely: Walker field $h_\textsc{w}^b = h_\textsc{w}^0 +\eta d/ \left(\sqrt{2}d_0\right)$ for a DW with $\mathcal{C}=+1$, Fig.~\ref{fig:vel_bDMI}(b) and chiral field $h_\textsc{c}^b = h_\textsc{w}^0 -\eta d/ \left(\sqrt{2}d_0\right)$ for a DW with $\mathcal{C}=-1$,  Fig.~\ref{fig:vel_bDMI}(c). The chiral field acquires two distinctive features: (i) at $h<h_\textsc{c}^b$ the ``unfavorable'' DW helicity $\mathcal{C} = -1$ has lower average velocity than for the opposite helicity, see Fig.~\ref{fig:vel_bDMI}(a); (ii) at $h_\textsc{c}^b<h<h_\textsc{w}^b$ the average velocity is the same for both initial helicities because after transient stage only one helicity survives, i.e. we have the traveling-wave motion with a single flip of the phase with $\varPhi^b\to\varPhi^b+\pi$. Movies in the Supplemental Materials~\cite{Note2} capture the dynamics of DWs with different helicities. 
During the transient stage, marked in Fig.~\ref{fig:vel_bDMI}(d) with black dashdotted lines, the DW with ``unfavorable'' helicity slows down during the phase flip and subsequently moves at a constant velocity. The DW helicity switching mechanism, illustrated by snapshots of simulated configuration in Fig.~\ref{fig:vel_bDMI}(f), occurs through the nucleation of vortex-antivortex pairs [Fig.~\ref{fig:vel_bDMI}(e) and the middle snapshot in Fig.~\ref{fig:vel_bDMI}(f)], followed by their annihilation, akin to the process described for vortex DWs in~\cite{Yan12}.  The origin of the vortex-antivortex pair generation is the nonlinear resonance of magnon modes with nonlinear coupling, the process originally studied for the planar vortex dynamics~\cite{Waeyenberge06,Gaididei10b}. Finally, at $h>h_\textsc{w}^b$ we see a sudden drop in the average DW velocity, see Fig.~\ref{fig:vel_bDMI}(a), which is typical for Walker breakdown. Here, both helicities periodically transform into each other as they propagate, showing periodic oscillations over the tube axis, so that the resulting average velocity is independent of the helicity.

The chirality-assisted DW behavior shown in Fig.~\ref{fig:vel_bDMI} resembles the dynamics of head-to-head~(tail-to-tail) vortex DWs in magnetic nanotubes studied in~\cite{Landeros10,Otalora12a,Yan12} and reflects geometry-induced chirality symmetry breaking. However, unlike the referenced works, which focus on DMI-free systems where chiral symmetry breaking originates nonlocally from magnetostatic interactions, here we observe a local chiral effect, arising from the interplay between bulk-type DMI and curvilinear geometry.

\subsubsection{The case of an axial magnetic field}\label{sec:bDMI_aField}

Next, we study the dynamics of DWs in a nanotube with an external magnetic field oriented along the tube axis, i.e. $\vec{h} = h \vec{e}_2$, or, in the rotated frame of reference, $\vec{h}^\psi = h \left(\sin\psi \vec{e}_1^\psi + \cos\psi \vec{e}_2^\psi\right)$. Note that here we consider magnetic fields below the Walker field~($h\ll h_\textsc{w}$). By substitution of ansatz~\eqref{eq:b_dw_solution} into the Zeeman energy and integration over $\xi_1$ and $\xi_2$ we obtain
\begin{equation}\label{eq:b_zeeman_axial}
	\frac{\mathcal{E}_\textsc{z}^b}{2\pi/\varkappa} = -2h\left(2 p q\sin\psi + \pi\Delta \cos\varPhi\cos\psi \right).
\end{equation}
Here, similarly to the case of the vortex field, we also have one term that pushes the DW ($\propto h\sin\psi$) and one that deforms the DW profile ($\propto h\cos\psi$). In this case, we restrict ourselves to the case of weak fields $h\ll h_\textsc{w}^0$, in order to avoid dynamical deformations of the DW profile.

\begin{figure}[t]
\includegraphics[width=\columnwidth]{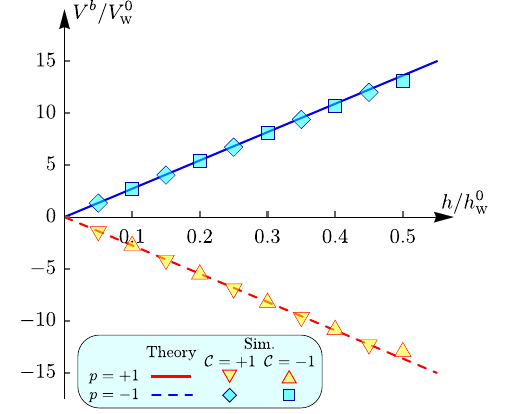}
\caption{\textbf{DW motion in an axial magnetic field in nanotubes with bulk-type DMI.} DW velocity $V^b$ is plotted as a function of applied axial magnetic field $\vec{h} = h \vec{e}_2$. Velocity is measured in units of DMI-free Walker velocity $V_\textsc{w}$, and the magnetic field is given in units of DMI-free Walker field, $h^0_\textsc{w}$. Lines correspond to the traveling-wave regime of motion~\eqref{eq:bDW-vel-A} for DWs of different topological charge, $p=\pm 1$, while symbols correspond to the results of numerical simulations (Appendix~\ref{app:sim}), where we use $p=\pm 1$, $d=0.25$, $\varkappa = 0.2$, $\eta = 0.01$, and $\varepsilon = 0.5$.\label{fig:vel_bDMI_ax}}
\end{figure}

In terms of the collective variables, the equations of motion take the form (for details see Appendix~\ref{app:bDW_exat})
\begin{equation}\label{eq:b_motion_axial}
	\begin{split}
		\frac{\eta}{\Delta}\dot{q} + p\,\dot{\varPhi} = 2ph\sin\psi,&\\
		p\dot{q} - \eta\Delta\dot{\varPhi} = 2p\,\frac{\mathcal{D}_{13}^{b(2)}}{d_0}&\cos\varPhi+\mathcal{K}_3^b\Delta\sin2\varPhi\\
		+&h\pi\Delta\sin\varPhi\cos\psi.
	\end{split}
\end{equation}

This set of equations has a traveling-wave regime with $\dot{q}=V^b=\text{const}$ and $\dot{\varPhi^b} = 0$:
\begin{subequations} \label{eq:b_trav_axial}
	\begin{align} \label{eq:bDW-vel-A}
		V^b =\, &2p\frac{h\Delta_0^b}{\eta}\sin\psi\approx -p\frac{d\varkappa}{\eta}h,\\
		\label{eq:bDW-phase-A}
		\varPhi^b \approx\, &\varPhi_0^b + \frac{hd}{2\varepsilon\left(2\mathcal{C}h+\varepsilon d_0\right)}\left(p\,\pi-\frac{d_0\varepsilon\varkappa}{\eta}\right).
	\end{align}
\end{subequations}
From~\eqref{eq:b_trav_axial} we can see that a DW has a finite velocity, and the direction of the DW motion is defined by the sign of product of the DW topological charge, $p$, and DMI strength, $d$, see Fig.~\ref{fig:vel_bDMI_ax}. The finite velocity \eqref{eq:bDW-vel-A} in an axial magnetic field is possible by the joint action of bulk-type DMI and curvature. In this case, the magnetization in the equilibrium state deviates from the easy-axis direction, and $\vec{h}\cdot\vec{m}_\text{tv} = Ch\sin\psi \ne 0$.

\section{Domain wall on a nanotube with interfacial-type DMI}
\label{sec:i_dmi}

\begin{figure}[t]
	\includegraphics[width=\columnwidth]{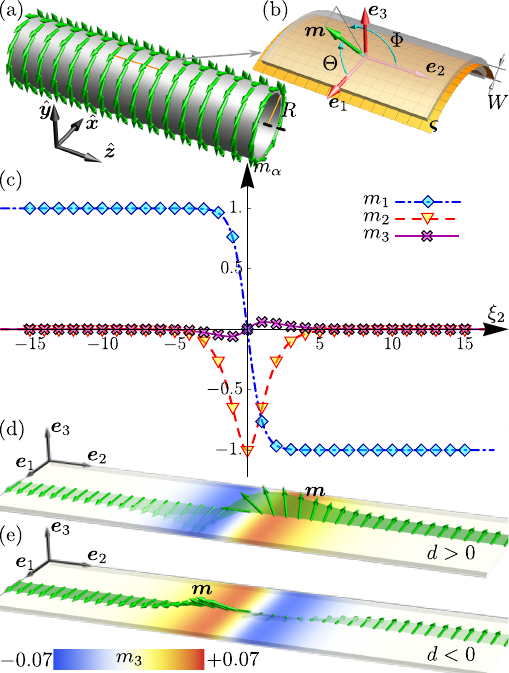}
	\caption{\textbf{Static domain walls in a cylindrical shell with interfacial-type DMI.} 
    (a),(b) Model of a cylindrical nanotube of mean radius $R$ and thickness $W$ with a magnetic texture.  Magnetization unit vector $\vec{m}$ (green arrows) forms a vortex state at equilibrium (a) and is parametrized by angles $\left\{\Theta,\Phi\right\}$ in the curvilinear reference frame (b). (c) Numerical simulation results (symbols) for magnetization components $m_\alpha$. Lines correspond to the analytical ansatz~\eqref{eq:i_dw_solution} with equilibrium DW profile parameters~\eqref{eq:i_equilibrium}. 
    (d) Reconstructed magnetization distribution from (c) in an unrolled cylindrical shell with $d>0$. Color scheme corresponds to the normal to the surface magnetization component $m_3$. Green arrows show the DW texture $\vec{m}(\xi_2)$, in which $m_3$ is amplified (10 times) for illustrative purpose due to its small magnitude. In numerical simulations we set $p=+1$, $d = 0.25$, $\varepsilon = 0.5$, and $\varkappa = 0.2$. (e) DW profile with the opposite sign of DMI constant, $d < 0$.
    \label{fig:iDW}}
\end{figure}

In contrast to the case of bulk-type DMI discussed in Sec.~\ref{sec:bDMI}, the symmetry of interfacial-type DMI, $\mathscr{E}_\textsc{d}^i=- \vec{m}\cdot \left[\left(\vec{e}_3\times\vec{\nabla}\right) \times \vec{m}\right]$, follows the in-surface easy direction $\vec{e}_1$, therefore the equilibrium magnetization texture forms one of two possible vortex states, $\vec{m}_\text{v} = C\vec{e}_1$ with $C=\pm1$, see Fig.~\ref{fig:iDW}(a). In this case, the unit vector of magnetization can be parametrized with angular variables ($\Theta$, $\Phi$) as shown in~Fig.~\ref{fig:iDW}(b):
\begin{equation}\label{eq:i_magn}
	\vec{m} = \cos\Theta\vec{e}_1+\sin\Theta\left(\cos\Phi\vec{e}_2+\sin\Phi\vec{e}_3\right),
\end{equation}
where $\Theta$ and $\Phi$ depend on the spatial and temporal coordinates.

The energy density for a nanotube with interfacial DMI can be written in the form, which is similar to the case of the bulk-type DMI~\eqref{eq:b_energy_rotated},
\begin{equation}\label{eq:i_energy_total}
	\begin{split}
		\mathscr{E}^i = \partial_\alpha m_\beta\partial_\alpha m_\beta &-\mathcal{K}^i_1 m_1^2 + \mathcal{K}^i_3m_3^2\\
		&+ \mathcal{D}_{\beta 3}^{i(\alpha)}\left(m_\beta\partial_\alpha m_3- m_3\partial_\alpha m_\beta\right),
	\end{split}
\end{equation}
with effective anisotropy and mesoscale DMI constants
\begin{equation}\label{eq:i_const}
	\begin{split}
		&\mathcal{K}_1^i = 1 - \varkappa^2,\quad \mathcal{K}_3^{i} = \varepsilon+\varkappa\left(\varkappa + d\right),\\
		&\mathcal{D}_{1 3}^{i(1)} = -\left(d+2\varkappa\right),\quad\mathcal{D}_{1 3}^{i(2)} = 0,\\
		&\mathcal{D}_{2 3}^{i(1)} = 0,\quad \mathcal{D}_{2 3}^{i(2)} = -d.\\
	\end{split}
\end{equation}
Here, we have only two non-zero mesoscale DMI terms due to similar symmetry of both intrinsic interfacial and geometry-governed DMI.

We take into account the nonlocal magnetostatic energy terms in the same way as we did in Sec.~\ref{sec:bDMI} for the case of bulk-type DMI. We directly include the local part coming from the surface energy $\mathcal{E}_\textsc{ms}^{\sigma-\sigma}$ by shifting the anisotropy ratio, $\varepsilon=\epsilon+Q^{-1}$, which appears in the effective anisotropy constant $\mathcal{K}_3^{i}$ in~\eqref{eq:i_const}. The contributions from nonlocal magnetostatic terms potentially allowing for chiral effects, $\mathcal{E}_\textsc{ms}^{\sigma-\rho}$ and $\mathcal{E}_\textsc{ms}^{g-\rho}$, we discuss further after the derivation of variational DW ansatz. 

\subsection{Static domain wall solution}

The structure of a DW in a nanotube with interfacial-type DMI between two vortices with opposite circulations $\cos\Theta_{\text{v}}=C=\pm1$ with opposite circulations can be described analytically as (see Appendix~\ref{app:iDW_sol}):
\begin{equation}\label{eq:i_dw_solution}
	\cos\Theta_\textsc{dw}^i = -p \tanh\frac{\xi_2}{\Delta},\quad \Phi_\textsc{dw}^i = \varPhi_0 + a\frac{\xi_2}{\Delta},
\end{equation}
where $a$ is a phase slope, and $\Delta$ is a DW width.

To study the finite curvature effects on the magnetization distribution in DWs, we apply a variational approach by using~\eqref{eq:i_dw_solution} as a DW Ansatz with the DW width $\Delta$, initial phase $\varPhi_0$, and phase slope $a$ being the variational parameters. By inserting Eq.~\eqref{eq:i_dw_solution} into the energy density functional~\eqref{eq:i_energy_total} and integrating over the $\xi_1$ and $\xi_2$, we obtain (see Appendices~\ref{app:iDW_sol} and \ref{app:iDW_ms}):
\begin{equation}\label{eq:i_dw_energy}
	\begin{split}
        \frac{\mathcal{E}_\textsc{dw}^i}{2\pi/\varkappa} = &\frac{2}{\Delta}\left(1+a^2\right)+2a\mathcal{D}_{23}^{i(2)}-a \Delta \frac{\pi w \varkappa }{4Q}\\
		&+\Delta\left[2\mathcal{K}_1^i+\mathcal{K}_{3}^i\left(1-\pi a\frac{\cos2\varPhi}{\sinh\left(\pi a\right)}\right)\right].
	\end{split}
\end{equation}

The minimization of~\eqref{eq:i_dw_energy} with respect to the DW width, phase, and its slope results in the following equilibrium values (approximation of a small phase slope)
\begin{equation}\label{eq:i_equilibrium}
	\begin{split}
		&\Delta^i_0  \approx 1/\sqrt{\mathcal{K}_1^i},\quad \cos\varPhi^i_0 = \mathcal{C}=\pm1,\\
		&a_0^i \approx -\frac{8\mathcal{D}_{23}^{i(2)}\sqrt{\mathcal{K}_1^i}-\pi w \varkappa/Q}{16 \left(\mathcal{K}_1^i+c\mathcal{K}_{3}^i\right)} ,\quad c = \pi^2 / 12.
	\end{split}
\end{equation}
In the limit case of small curvature and DMI strength, $\Delta^i_0 \approx 1 + \varkappa^2/2$ and $a_0^i \approx \left[d+\pi w \varkappa/\left(8Q\right)\right]/\left(2+2c\varepsilon\right)$. The comparison of these predictions with numerical simulations confirms our analytical calculations, see Fig.~\ref{fig:iDW}.

Note that in the case of interfacial-type DMI, the magnetostatic contribution manifests not only as a shift in the anisotropy constant but also through the terms $\propto w \varkappa /Q$ in~\eqref{eq:i_dw_energy} and~\eqref{eq:i_equilibrium}. These terms, arising from the interactions ${E}_\textsc{ms}^{g-\rho}$ and ${E}_\textsc{ms}^{\sigma-\rho}$, lead to the emergence of a DW configuration with a favorable sign of phase slope $a_0^i$, determined by the interplay between local and non-local contributions (see details in Appendix~\ref{app:iDW_ms}).
Importantly, the DW topological charge $p = \pm 1$ has no impact on the magnetostatic energy in the studied system, meaning chiral effects are absent. The preferred phase slope $a_0^i$ remains the same regardless of $p$ or the helicity $\mathcal{C}$. 

\subsection{Domain wall dynamics \\ in an external magnetic field}

\begin{figure}[t]
	\includegraphics[width=\columnwidth]{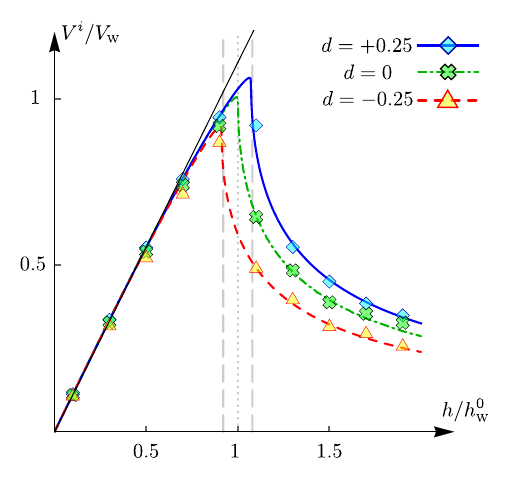}
	\caption{\textbf{Domain wall motion in nanotubes with interfacial-type DMI in the applied vortex magnetic field $\vec{h} = h \vec{e}_1$}. DW velocity $V^i$ is measured with respect to DMI-free Walker velocity $V_\textsc{w}$, and the magnetic field value is given in units of DMI-free Walker field $h_\textsc{w}^0$ (dotted vertical gray line). The solid black line illustrates the traveling wave regime~\eqref{eq:i_trav_vortex}. Colored lines show the numerical solution of the equations of motion~\eqref{eq:i_motion_vortex}. Dashed vertical gray lines mark the predicted values of the Walker field~\eqref{eq:i_walker_field} for different signs of DMI strength $d$. Symbols correspond to the results of numerical simulations (Appendix~\ref{app:sim}) with $p=+1$, $d=\pm0.25$, $\varkappa = 0.2$, $\eta = 0.01$, and $\varepsilon = 0.5$.\label{fig:vel_iDMI}}
\end{figure}

In order to derive effective equations of the DW motion we use the collective variable approach based on a generalized $q-\Phi$ model~\cite{Kravchuk14}
\begin{equation} \label{eq:i_q-phi}
\cos\Theta_\textsc{dw}^i = -p \tanh\frac{\xi_2-q(t)}{\Delta},\quad \Phi_\textsc{dw}^i = \varPhi(t)+a \frac{\xi_2-q(t)}{\Delta}.
\end{equation}

For the case of interfacial-type DMI, we consider the dynamics of DWs in a nanotube only in a vortex magnetic field, i.e. $\vec{h}=h\vec{e}_1$. For the case of $\vec{h}\|\vec{e}_2$, DWs are \textit{immobile} since $\vec{h}\cdot\vec{m}=0$ in the ground state. By substitution of the ansatz~\eqref{eq:i_q-phi} into the Zeeman energy and integration over $\xi_1$ and $\xi_2$ coordinates, we obtain the energy term which corresponds to the Zeeman energy density~\eqref{eq:b_zeeman_vortex} with $\psi=0$.

In terms of the collective variables, the equations of motion take a form
\begin{equation}\label{eq:i_motion_vortex}
	\begin{split}
		\eta\frac{\dot{q}}{\Delta}+ \left(p-a\eta\right)\dot{\varPhi} &= 2ph,\\
		\left(p+a\eta\right)\dot{q}-\eta\Delta\dot{\varPhi} &= \mathcal{K}_{3}^i\Delta\sin2\varPhi.
	\end{split}
\end{equation}
The DW width, $\Delta$, and phase slope, $a$, are assumed to be the slave variables, i.e., $\Delta^i\left[\varPhi^i(t)\right] = 1/\sqrt{\mathcal{K}_1^i+\mathcal{K}_{3}^i\sin^2\varPhi^i}$ and $a^i\left[\varPhi^i(t)\right] = \left(\frac{\pi w \varkappa }{8Q}\Delta^i-\mathcal{D}_{23}^{i(2)}\right)\Delta^i/\left[2+2c\left(\Delta^i\right)^2\mathcal{K}_{3}^i\cos2\varPhi^i\right]$. The behavior of the DW width and phase slope is discussed in detail in Appendix~\ref{app:iDW_exat}.

The set of equations of motion~\eqref{eq:i_motion_vortex} has a traveling wave solution with:
\begin{equation}\label{eq:i_trav_vortex}
	\begin{split}
		V^i = 2p\frac{h\Delta^i_0}{\eta},\qquad&\varPhi^i =\varPhi_0^i +\frac{p}{2}\arcsin\frac{h}{h_\textsc{w}^i},\\
	\end{split}
\end{equation}
with the Walker field:
\begin{equation}\label{eq:i_walker_field}
	h_\textsc{w}^i \approx h_\textsc{w}^0 + d\frac{\eta}{2}\left[\varkappa + p\frac{\varepsilon\eta}{2\sqrt{1+\varepsilon/2}}\right]+p\frac{\varepsilon  \eta^2 \varkappa}{2+\varepsilon}\frac{\pi w}{16Q}.
\end{equation}

Thus, the Walker field for a DW in a nanotube with interfacial-type DMI exhibits a linear shift with respect to the DMI strength $d$, as seen from the second term on the right-hand side of~\eqref{eq:i_walker_field}, which arises from local interactions. While it contains the DW topological charge $p$, reasonable values of the damping constant $\eta \sim 10^{-2}$ render the $p$-term negligibly small compared to the curvature $\varkappa$. The same reasoning applies to the magnetostatics-generated last term in~\eqref{eq:i_walker_field}.
We conclude that the DW motion does not exhibit chiral effects in the considered setup, and the shift in the Walker field depends solely on the sign of the DMI strength (see Fig.~\ref{fig:vel_iDMI}). Moreover, the Walker field shift is weaker here compared to the case with bulk-type DMI (Sec.~\ref{sec:bDMI_vField}), being quadratic with respect to small parameters $\propto \varkappa d$, while for the bulk-type DMI, the leading term in the field shift~\eqref{eq:b_walker_field} scales linearly on the order of $d$.

\section{Conclusions}\label{sec:conclusions}

The effects of DMI and curvature on the statics and field-driven DW motion are analyzed using cylindrical biaxial magnetic nanotubes as a model system. The results demonstrate that the presence of DMI leads to specific deformations in the DW profile, with the pattern and magnitude of these distortions dictated by the DMI type and strength, respectively. In particular, bulk-type DMI generates a symmetric normal (out-of-surface) magnetization component (Fig.~\ref{fig:bDW}), whereas interfacial-type DMI induces asymmetric distortions in the normal magnetization component (Fig.~\ref{fig:iDW}). Note that we did not observe the tilt (with respect to the nanotube's axis) of the DW profile reported in the pre-print by Josten~\textit{et al.}~\cite{Josten21}. Both analytical calculations and numerical (spin-lattice and full-scale micromagnetic) simulations predict DWs perpendicular to the nanotube's axis. 

The joint action of curvature and bulk-type DMI results in a tilting of the vortex ground state of magnetization by an angle $\psi\propto \varkappa d$ as compared to the DMI-free case. This effect enables the translational motion of a DW with constant velocity $V^b/h\propto \varkappa d$ in an axial magnetic field. Moreover, in the case of a vortex magnetic field,  bulk-type DMI leads to the appearance of the preferable combination of DW topological charge and helicity, $\sign(p\,\mathcal{C}d)<0$. The latter effect induces chirality breaking in DW motion: a DW with non-preferable chirality undergoes a vortex-antivortex pair mediated breakdown before settling into motion at a constant velocity, as shown in Fig.~\ref{fig:vel_bDMI}(d)-(f). Additionally, bulk-type DMI increases the Walker field proportionally to the DMI constant, thereby extending the range of applied fields that enable steady DW motion.

In systems with interfacial-type DMI, the Walker field also exhibits a linear shift with respect to the DMI strength; however, this effect is weaker than in the bulk-type DMI case, and no chiral breakdown in DW motion is observed. Notably, the asymmetry of the static DW profile in this case is jointly tuned by DMI and magnetostatics, where the latter contributes in the thin-film limit to both effective anisotropy and DMI constants. Nonlocal interactions, however, do not generate chiral effects in the DW dynamics and eventually result in a minor additional shift in the Walker field.

The synergy between symmetry and geometry in magnetic nanotubes provides an enhanced toolkit for exploring topological effects and features of curvilinear magnetization dynamics. Our predictions pave the way for tailoring the properties of DW textures in thin cylindrical nanotubes, enabling their use as critical components in nanoelectronic devices.

\section*{Acknowledgments}

We thank Ulrike Nitzsche for technical support. This work was financed in part via the German Research Foundation (DFG) Grants No. YE 232/2-1 and MC 9/22-1. S.K. acknowledges the support from the Alexander von Humboldt Foundation.

\appendix

\section{Equilibrium magnetization texture}
\label{app:param}

To identify possible equilibrium magnetization textures in DMI-free nanotubes, we introduce the following parametrization of the unit magnetization: $\vec{m} = \vec{e}_1 \cos\varphi \sin\vartheta + \vec{e}_2\cos\vartheta+\vec{e}_3\sin\varphi \sin\vartheta$, where $\{\vartheta,\varphi\} = \{\vartheta(\xi_1),\varphi(\xi_2)\}$. The equilibrium values of $\vartheta$ and $\varphi$ are determined by the equations $\delta\mathcal{E}/\delta\vartheta = 0$ and $\delta\mathcal{E}/\delta\varphi = 0$. By solving these equations we find homogeneous and periodic equilibrium states.

The first one is the vortex state. This state corresponds to the flux-free magnetization distribution, in which magnetization is oriented tangentially to the cylindrical surface, with $\vec{m}=\pm\vec{e}_1$, i.e. $\vartheta = \pi/2$ and $\cos\varphi = \pm1$. The energy of the vortex state is $\mathcal{E}^\text{v} = \varkappa^2$.

\begin{figure}[t]
	\includegraphics[width=0.85\columnwidth]{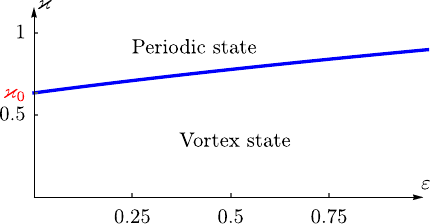}
	\caption{\textbf{Phase diagram of equilibrium magnetization textures in a DMI-free magnetic nanotube}. The parameter space, curvature $\varkappa$ -- easy-surface anisotropy $\varepsilon$, hosts two possible equilibrium configurations, periodic state and vortex state. The boundary curve is plotted with prediction~\eqref{eq:crit_kappa}.\label{fig:eq_states}}
\end{figure}

Another possible equilibrium configuration is the periodic magnetization distribution with $\vartheta = \pi/2$ and $\varphi$ defined by a pendulum equation,
\begin{equation}
    \partial_{11}\varphi - \lambda\sin\varphi\cos\varphi=0, \quad \lambda = 1+\varepsilon,
\end{equation}
with a solution,
\begin{equation}\label{eq:periodic_sol}
		\varphi = \text{am}\left[\sqrt{\mathfrak{C}}\xi_1,-\frac{\lambda}{\mathfrak{C}}\right].
\end{equation}
Here, $\mathfrak{C}$ is an integration constant. The solution \eqref{eq:periodic_sol} has a period
\begin{equation}
	T = \frac{4}{\sqrt{\mathfrak{C}}}\mathrm{K}\left[-\frac{\lambda}{\mathfrak{C}}\right] = \frac{2\pi}{q\varkappa},
\end{equation}
which is also predefined by geometry. Here $q$ defines $2q$ number of domain walls on the cylindrical surface. The energy of the periodic solution is defined as
\begin{equation}
	\mathcal{E}^\text{p} = \mathcal{E}^\text{v} -\mathfrak{C} - 2q\varkappa \left[\varkappa - \frac{2\sqrt{\mathfrak{C}}}{\pi}\mathrm{E}\left(-\frac{\lambda}{\mathfrak{C}}\right)\right],
\end{equation}
where $\mathrm{E}[x]$ is the complete elliptic integral of the second kind~\cite{Abramowitz72}. For a planar film, the transition between the homogeneous and periodical state is characterized by an infinite increase in the period of the spiral state~\cite{Rohart13}. Although for the cylindrical surface the period is finite in the transition point, for the limit case $\mathfrak{C}\to 0$ one has $T\to\infty$. Using that $\mathfrak{C}\to 0$ in this limit, we obtain from the equality  $\mathcal{E}^\text{p} = \mathcal{E}^\text{v}$ the analytical expression for the critical curvature value $\varkappa_c$:
\begin{equation}\label{eq:crit_kappa}
	\varkappa_c = \frac{2\sqrt{1+\varepsilon}}{\pi}.
\end{equation}
In the limit case of vanishing easy-surface anisotropy~($\varepsilon=0$) the critical curvature is defined as $\varkappa_0=2/\pi$ which is the same as for the nanotubes with easy-normal anisotropy~\cite{Yershov20} and analogous to the effect of spontaneous formation of the onion state in nanorings when curvature exceeds some critical value~\cite{Sheka15}. The phase diagram of equilibrium states is shown in Fig.~\ref{fig:eq_states}.

\section{Magnetic nanotube with DMI of bulk type}\label{app:bDMI}

\subsection{Micromagnetic energy in a $\psi$-frame}
\label{app:bDMI_psi_frame}

For the case of bulk-type DMI, one can obtain from the energy~\eqref{eq:model} a term which is quadratic with respect to the magnetization components, i.e. so-called effective anisotropy in the form $\mathscr{E}_\textsc{a}^{b\text{(eff)}} = \mathcal{K}_{\alpha\beta} m_\alpha m_\beta$ where $\mathcal{K}_{\alpha\beta}$ is the effective total anisotropy matrix with coefficients
\begin{equation}\label{eq:b_anisotropy}
	\mathcal{K}_{\alpha\beta} =\left(\begin{matrix}
		\varkappa^2-1 & d\varkappa/2&0\\
		d\varkappa/2&0&0\\
		0&0&\varepsilon+\varkappa^2\\
	\end{matrix}\right).
\end{equation}
This matrix has non-diagonal components. This means that the homogeneous magnetization structure is not oriented along the curvilinear basis, i.e. magnetization deviates from $\vec{e}_1$ direction by an angle $\psi \approx -d\varkappa / 2$ (direction is defined by the sign of DMI constant $d$). One can easily diagonalize $\mathcal{K}_{\alpha\beta}$, by using a unitary transformation (rotation in a local tangential plane) of the vector~$\vec{m}$
\begin{equation}
	\begin{split}
		\vec{m} = U\vec{\mu},&\quad \vec{\mu} = U^{-1}\vec{m},\quad \vec{\mu} = \left(\mu_1,\mu_2,\mu_3\right)^\textsc{t},\\
		&U=\left(\begin{matrix}
			\cos\psi & -\sin\psi& 0\\
			\sin\psi & \cos\psi & 0\\
			0 & 0& 1
		\end{matrix}\right).
	\end{split}
\end{equation}

By choosing $\psi = \arctan\left[\left(1-\varkappa^2-\mathcal{K}^b_1\right)/\left(d\varkappa\right)\right]$ as the rotation angle, where $\mathcal{K}^b_1=\sqrt{\left(1-\varkappa^2\right)^2+d^2\varkappa^2}$, it is possible to reduce the anisotropy energy \eqref{eq:b_anisotropy} to the form
\begin{equation*}
	\begin{split}
		\mathscr{E}_\textsc{a}^{b\text{(eff)}} &= -\mathcal{K}^b_1 \mu_1^2 + \mathcal{K}^b_3 \mu_3^2,\quad \mathcal{K}^b_1\approx 1- \varkappa^2\left(1-\frac{d^2}{2}\right),\\
		\mathcal{K}^b_3 &=\varepsilon+\frac{1+\varkappa^2-\mathcal{K}^b_1}{2}	\approx \varepsilon + \varkappa^2\left(1-\frac{d^2}{4}\right).
	\end{split}
\end{equation*}
Here, coefficient $\mathcal{K}^b_1$ characterizes the strength of the effective easy-axis anisotropy, while $\mathcal{K}^b_3$ gives the strength of the effective easy-surface anisotropy. The direction of the effective easy axis is determined by $\vec{e}^\psi_1$ and the hard axis by $\vec{e}^\psi_3$:
\begin{equation}
	\begin{split}
		\vec{e}_1^\psi &= \vec{e}_1\cos\psi + \vec{e}_2\sin\psi,\\
		\vec{e}_2^\psi &= -\vec{e}_1\sin\psi + \vec{e}_2\cos\psi,\\
		\vec{e}_3^\psi &= \vec{n}=\vec{e}_3.
	\end{split}
\end{equation}
Note that for any finite $\psi$ the effective anisotropy direction $\vec{e}_1^\psi$ deviates from the magnetic anisotropy direction $\vec{e}_1$. This deviation vanishes in tubes with zero DMI ($d\to0$).

Apart from the effective anisotropy, curvature shows up in the effective DMI. In the new frame of reference~(rotated $\psi$-frame), the mesoscale DMI energy reads
\begin{equation}
	\begin{split}
		&\mathscr{E}_\textsc{d}^{b\text{(eff)}} = \mathcal{D}_{\beta 3}^{b(\alpha)}\left(\mu_\beta\partial_\alpha\mu_3-\mu_3\partial_\alpha\mu_\beta\right),\\
		&\mathcal{D}_{1 3}^{b(1)} = -2 \varkappa\cos\psi - d \sin\psi\approx -2\varkappa+\frac{d^2}{2}\varkappa,\\
		&\mathcal{D}_{2 3}^{b(1)} = -d\cos\psi +2\varkappa \sin\psi\approx -d - d\varkappa^2,\\
		&\mathcal{D}_{1 3}^{b(2)} = d\cos\psi\approx d,\quad \mathcal{D}_{2 3}^{b(2)} = -d\sin\psi \approx \frac{d^2}{2}\varkappa.\\
	\end{split}
\end{equation}
With this, we get to the energy density in the form of~\eqref{eq:b_energy_rotated}.

\subsection{Static DW solution}
\label{app:bDW_sol}

We consider the no-driving case $h=0$, in which case the ground state is defined by the set of static equations $\delta\mathcal{E}^b/\delta\theta=0$ and $\delta\mathcal{E}^b/\delta\phi=0$, which read:

\begin{equation}\label{eq:bDMI_euler}
	\begin{aligned}
		&\partial_{11}\theta+\partial_{22}\theta + \cos\phi\sin^2\theta \left[\mathcal{D}_{13}^{b(1)}\partial_1\phi+\mathcal{D}_{13}^{b(2)}\partial_2\phi\right]\\
		&- \sin\theta\cos\theta \bigl[\left(\partial_1\phi\right)^2+\left(\partial_2\phi\right)^2\mathcal{K}_1^b+\mathcal{K}_3^b\sin^2\phi\\
        &+ \mathcal{D}_{23}^{b(1)} \partial_1\phi + \mathcal{D}_{23}^{b(2)} \partial_2\phi\bigr]= 0,\\
		&\partial_1\left(\sin^2\theta\partial_1\phi\right) + \partial_2\left(\sin^2\theta\partial_2\phi\right)  \\
        &+ \sin\theta \cos\theta \left[ \mathcal{D}_{23}^{b(1)} \partial_1\theta + \mathcal{D}_{23}^{b(2)} \partial_2\theta\right]\\
		&- \cos\phi \sin^2\theta \left[\mathcal{K}_3^b \sin\phi + \mathcal{D}_{13}^{b(1)} \partial_1\theta + \mathcal{D}_{13}^{b(2)} \partial_2\theta\right] = 0.
	\end{aligned}
\end{equation}

Without DMI ($d=0$) the vortex ground state of the system with energy~\eqref{eq:model} is doubly degenerated: $\theta=0$ and $\theta=\pi$. The transition between domains of different ground states forms a DW. Structure of the DW can be found as a solution of~\eqref{eq:bDMI_euler} with boundary conditions $\cos\theta\left(\pm\infty\right)=\mp p$. This solution is well known, 
\begin{equation}\label{eq:dw_dmi_free}
	\theta_\textsc{dw}^0\left(\xi_2\right) = 2\arctan e^{p \xi_2/\Delta}, \quad \cos\phi_\textsc{dw}^0 = \pm 1,
\end{equation}
where $\Delta = 1/\sqrt{\mathcal{K}_1^b} = 1/\sqrt{1-\varkappa^2}$ is the width of a static DW. Nevertheless, the DW solution~\eqref{eq:dw_dmi_free} does not satisfy Eqs.~\eqref{eq:bDMI_euler} in case $d\neq 0$. In the following  we introduce the deformation of the DW solution~\eqref{eq:dw_dmi_free} induced by the DMI, considering DMI as a small perturbation. For this purpose we introduce small deviations from the non-perturbed solution~\eqref{eq:dw_dmi_free}
\begin{equation}\label{eq:dw_deformed}
	\theta=\theta_\textsc{dw}^0 + \vartheta, \quad \phi=\phi_\textsc{dw}^0 +\frac{\varphi}{\sin\theta_\textsc{dw}^0(\xi_2)}.
\end{equation}
Substituting now~\eqref{eq:dw_deformed} into~\eqref{eq:bDMI_euler} and linearizing the obtained equation with respect to the deviations one obtains the following
equations for the deviations $\vartheta$ and $\varphi$:
\begin{subequations} \label{eq:dw_b_deformed}
\begin{align} \label{eq:dw_b_deformed-vartheta}
&\vartheta'' + \left(\frac{2}{\cosh^2\zeta}-1\right)\vartheta=0,\\
\label{eq:dw_b_deformed-varphi}
&\varphi'' + \left(\frac{2}{\cosh^2\zeta}-\alpha^2 \right)\varphi = \frac{\beta^b}{\cosh^2\zeta},
\end{align}
\end{subequations}
where $\zeta = \xi_2/\Delta$ and $\partial_1\vartheta=\partial_{11}\vartheta = \partial_1\varphi =\partial_{11}\varphi=0$, $\alpha =\sqrt{1+ {\mathcal{K}_3^b}/{\mathcal{K}_1^b}}$, and $\beta^b = p\,\mathcal{C} \mathcal{D}_{13}^{b(2)}/{\sqrt{\mathcal{K}_1^b}}$. In~\eqref{eq:dw_b_deformed} prime denotes the derivative with respect to $\zeta$.

The non-trivial solution of the homogeneous Eq.~\eqref{eq:dw_b_deformed-vartheta} $\vartheta = \mathrm{d}\theta_\textsc{dw}^0/\mathrm{d}\zeta = 1/\cosh\zeta$ corresponds to the Goldstone mode of the DW~\eqref{eq:dw_dmi_free}. The function $\varphi(\zeta)$ is exponentially localized, $\varphi \propto \beta^b \exp(-\alpha\zeta)$ when $\zeta\to\infty$, and tends to $\beta^b/(2-\alpha^2)$ when $\zeta\to0$.

\subsection{Magnetostatic energy}\label{app:bDW_ms}

The most contribution to the magnetostatic energy is generated by the interaction of surface charges, induced by the magnetisation component normal to the tube's surface, $\sigma^{\pm}=\vec{m}\cdot\vec{n}^{\pm}=\pm\mu_3$ at surfaces $R^{\pm}=R\pm W/2$. The corresponding energy term $\mathcal{E}_\textsc{ms}^{\sigma-\sigma}$ is local in the thin-shell limit, scaling as shell thickness $W$. 
We can calculate this contribution as 
\begin{equation}
{E}_\textsc{ms}^{\sigma-\sigma}=\frac{\mu _{0}M_\text{s}}{2}\int  \left( \sigma^+
U_{\sigma }^+ +\sigma^-  U_{\sigma }
^-\right)\textrm{d}x_1\mathrm{d}x_2,
\label{eq:MSsurf}
\end{equation}
where $U_{\sigma }^\pm=U_{\sigma }
\left(x_3=R^\pm\right)$ is the potential of the surface charges given by:
\begin{equation*}
U_{\sigma }\left( \mathbf{x}\right) =\frac{M_\text{s}}{4\pi }\iint\limits_{x_{3}^{
\prime }=R^{\pm }}\frac{\sigma \left(
\mathbf{x}^{\prime
} \right) }{\left\vert \mathbf{x}-\mathbf{x}^{\prime
}\right\vert  } \mathrm{d}x_{1}^{\prime }\mathrm{d}x_{2}^{\prime }.
\label{eq:Usurf}
\end{equation*}
Following the approach in~\cite{Landeros10}, we expand the cylindrical Green function in terms of Bessel functions $J_N(x)$ and perform the integral over $x_{1}^{\prime }$, obtaining:
\begin{equation*}
	\begin{split}
		U_{\sigma }\left( x_2,x_{3}\right) 
		&=\frac{M_\text{s}}{2}\int_{0}^{\infty
}\mathrm{d}k\,J_{0}\left( kx_{3}\right) \left[ R^{+}J_{0}\left( kR^{+}\right) \right. \\
&- \left. R^{-}J_{0}\left( kR^{-}\right) \right] 
		\int_{-\infty }^{\infty
}\mathrm{d}x_{2}^{\prime }e^{-k\left\vert x_{2}-x_{2}^{\prime }\right\vert }\mu
_{3}^{+}\left( x_{2}^{\prime }\right).\\
	\end{split}
\end{equation*}
Next, using the integral representation $e^{-k\left\vert x_{2}-x_{2}^{\prime }\right\vert }=\frac{1}{2\pi }%
\int_{-\infty }^{\infty }\frac{2k}{k^{2}+q^{2}}e^{iq\left(
x_{2}-x_{2}^{\prime }\right) }\textrm{d}q$, substituting the resulting potential to~\eqref{eq:MSsurf}, and integrating over $x_1$ we arrive at
\begin{equation*}
	\begin{split}
		E_{\mathrm{MS}}^{\sigma -\sigma }&=\frac{1}{2}\mu _{0}M_\text{s}^{2}\int_{-\infty
}^{\infty }\textrm{d}q  \\
		& \times \int_{0}^{\infty }\frac{k \textrm{d}k}{k^{2}+q^{2}}\left[
R^{+}J_{0}\left( kR^{+}\right) -R^{-}J_{0}\left( kR^{-}\right) \right]
^{2} \\
		&\times \int_{-\infty }^{\infty }\textrm{d} x_{2}e^{iqx_{2}}\mu _{3}^{+}\left(
x_{2}\right) \int_{-\infty }^{\infty } \textrm{d}x_{2}^{\prime }e^{-iqx_{2}^{\prime
}}\mu _{3}^{+}\left( x_{2}^{\prime }\right). \\
	\end{split}
\end{equation*}

After having integrated over $k$ and expanded the result in powers of $W$ (thin-shell limit), we see that
\begin{equation}
E_{\mathrm{MS}}^{\sigma -\sigma } \simeq \frac{1}{2}\mu _{0}M_\text{s}^{2}RW\int_{-\infty
}^{\infty } \widetilde\mu _{3}^{+}\left(-q \right) \widetilde\mu _{3}^{+}\left(q \right) \textrm{d}q, 
\end{equation}
where $\widetilde\mu _{3}^{+}\left(q \right)$ is a Fourier transform of $\mu _{3}^{+}(x)$.
Applying Parseval's theorem, we get
\begin{equation}
E_{\mathrm{MS}}^{\sigma -\sigma } \simeq \frac{1}{2}\mu _{0}M_\text{s}^{2} \left(2\pi R \right) W\int_{-\infty
}^{\infty }dx_2 \left\vert \mu _{3}\left(x_2 \right)\right\vert^2, 
\label{eq:MSthin}
\end{equation}
which allows us to directly include this term to the energy density~\eqref{eq:b_energy_rotated} as $\mathcal{E}_\textsc{ms}^{\sigma-\sigma}=wQ E_{\mathrm{MS}}^{\sigma -\sigma }/E_0=w\iint \mu_3^2 \textrm{d}\xi_1\mathrm{d}\xi_2$ 
by shifting the anisotropy ratio as $\varepsilon=\epsilon+Q^{-1}$, which emerges then in the effective anisotropy constant $\mathcal{K}^b_3$~\eqref{eq:b_const_rotated}.

Using ansatz~\eqref{eq:b_dw_solution} for the case of bulk-type DMI, we substitute $\mu_3 = \sin\varPhi / \cosh(\xi_2/\Delta)$ and obtain $\mathcal{E}_\textsc{ms}^{\sigma-\sigma}= w(2\pi / \varkappa) 2\Delta \sin^2\varPhi$, as appears in~\eqref{eq:b_dw_energy} derived from~\eqref{eq:model}. 

The volume and geometric magnetostatic charges for the magnetization distribution given by~\eqref{eq:b_dw_solution} are as follows: 
\begin{equation*}
\rho ^{b} = \frac{1}{\Delta }
\frac{\tanh \left(\xi_2/\Delta \right)}
{\cosh \left(\xi_2/\Delta \right)}\cos \varPhi;  \quad
g^{b} = -\frac{1}{\xi_3}
\frac{\sin \varPhi}{\cosh \left(\xi_2/\Delta \right)}. 
\end{equation*}

The magnetic symmetry of terms corresponding to interactions $\mathcal{E}_\textsc{ms}^{\rho-\rho}$, $\mathcal{E}_\textsc{ms}^{g-g}$, and $\mathcal{E}_\textsc{ms}^{\sigma-g}$ restricts symmetry breaking and cannot lead to chiral effects~\cite{Sheka20}, resulting just in a correction of order $\sim$ $\lito(w)$~\cite{Carbou01,Kohn05a,Kohn05,Fratta20a} to the total energy. The remaining terms $\mathcal{E}_\textsc{ms}^{\sigma-\rho}$ and $\mathcal{E}_\textsc{ms}^{g-\rho}$, while generally allowing for nonlocal chiral effects, in our case conserve the chiral symmetry due to the same parity of
the components $\mu_2$ and $\mu_3$ with respect to $\xi_2\rightarrow -\xi_2$. 

We conclude that for the case of a DW in a nanotube with bulk-type DMI only local chiral effects are present. The account of the local magnetostatic contribution~\eqref{eq:MSthin} is sufficient for the considered range of geometrical and material parameters. 

\subsection{Exact form of the equations of motion for DW}\label{app:bDW_exat}

The equations of motion can be derived from the Lagrangian $\mathcal{L}$ and the dissipative function $\mathcal{R}$:
	\begin{eqnarray}
		&\mathcal{L} = - \iint \phi\dot{\theta}\sin\theta\mathrm{d}\xi_1\mathrm{d}\xi_2 - \mathcal{E},\\
		&\mathcal{R} = \frac{\eta}{2}\iint \left(\dot{\theta}^2+\dot{\phi}^2\sin^2\theta\right)\mathrm{d}\xi_1\mathrm{d}\xi_2,
        \label{eq:LR_functions}
	\end{eqnarray}
	where $\eta$ is the Gilbert damping parameter.

By substituting the ansatz~\eqref{eq:b_q-phi} into~\eqref{eq:LR_functions} one obtains
\begin{equation}\label{eq:b_LR}
	\begin{split}
		\frac{\mathcal{L}_\textsc{dw}^b}{2\pi/\varkappa} &= 2p\varPhi \dot{q} - \frac{\mathcal{E}_\textsc{dw}^b}{2\pi/\varkappa},\\ \frac{\mathcal{R}_\textsc{dw}^b}{2\pi/\varkappa} &= \frac{\eta}{\Delta}\left[\dot{q}^2+\dot{\varPhi}^2\Delta^2+c \dot{\Delta}^2\right],
	\end{split}
\end{equation}
with total energy in an external magnetic field
\begin{widetext}
	\begin{equation}\label{eq:b_energyZ}
		\begin{split}
			&\frac{\mathcal{E}_\textsc{dw}^b}{2\pi/\varkappa} = \frac{2}{\Delta}+2\Delta\left(\mathcal{K}^b_1\!+{\mathcal{K}}_3^b\sin^2\varPhi\right)+p\,\pi\mathcal{D}_{13}^{b(2)}\sin\varPhi+\frac{\mathcal{E}_\textsc{z}^b}{2\pi/\varkappa},\\
			&\frac{\mathcal{E}_\textsc{z}^b}{2\pi/\varkappa} = -4h\begin{cases}
				p\,q\cos\psi - \frac{\pi}{2}\Delta \cos\varPhi\sin\psi,\quad \vec{h}||\vec{e}_1\text{ -- vortex magnetic field},\\
				p\,q\sin\psi + \frac{\pi}{2}\Delta \cos\varPhi\cos\psi,\quad \vec{h}||\vec{e}_2\text{ -- axial magnetic field}.
			\end{cases}
		\end{split}
	\end{equation}
\end{widetext}

Lagrangian and dissipative function~\eqref{eq:b_LR} produce the following set of equations of motion:
\begin{eqnarray}\label{eq:bDW_equation_of_motion}
		\frac{\eta}{\Delta}\dot{q} + p \dot{\varPhi} &=& -\frac{\varkappa}{4\pi}\frac{\partial\mathcal{E}_\textsc{z}^b}{\partial q},\\
		p\dot{q} - \eta\Delta\dot{\varPhi} &=& 2p\,\frac{\mathcal{D}_{13}^{b(2)}}{d_0}\cos\varPhi+{{\mathcal{K}}_3^b}\Delta\sin2\varPhi+\frac{\varkappa}{4\pi}\frac{\partial\mathcal{E}_\textsc{z}^b}{\partial \varPhi},\nonumber \\
		c\eta \frac{\dot{\Delta}}{\Delta} &=&\frac{1}{\Delta^2}-\mathcal{K}_1^b-{{\mathcal{K}}_3^b}\sin^2\varPhi-\frac{\varkappa}{4\pi}\frac{\partial\mathcal{E}_\textsc{z}^b}{\partial \Delta}.\nonumber    
\end{eqnarray}

(i) For the case of a vortex magnetic field ($\vec{h}\|\vec{e}_1$), the third equation in~\eqref{eq:bDW_equation_of_motion} shows that $\Delta$ relaxes towards its equilibrium value $\Delta_0^b\approx1/\sqrt{\mathcal{K}_1^b+{{\mathcal{K}}_3^b}\sin^2\varPhi^b}$. The characteristic time of this relaxation is proportional to the damping $\propto \eta$~\cite{Hillebrands06}. Usually $\eta\ll1$, therefore one can conclude that the DW width is a slave variable $\Delta\left[\varPhi(t)\right]$, and DW dynamics can be described by the set~\eqref{eq:b_motion_vortex} with the equilibrium DW width $\Delta=\Delta^b_0$.

(ii) For the axial magnetic field ($\vec{h}\|\vec{e}_2$), the third equation in~\eqref{eq:bDW_equation_of_motion} shows that $\Delta$ relaxes to the value $\Delta_0^b=1/\sqrt{\mathcal{K}_1^b+{{\mathcal{K}}_3^b}\sin^2\varPhi^b-\pi h\cos\psi}$, i.e. external field results in a small deformation of DW shape. As mentioned in the Sec.~\ref{sec:bDMI_aField}, here we consider fields smaller than the Walker field, $h\ll h_\textsc{w}^0$. In this case, field-induced deformations are negligible, and the equilibrium DW width can be defined as $\Delta_0^b\approx1/\sqrt{\mathcal{K}_1^b+{{\mathcal{K}}_3^b}\sin^2\varPhi^b}$. Thus, we conclude that here the DW width is also a slave variable $\Delta\left[\varPhi(t)\right]$, and the DW dynamics can be described by the set~\eqref{eq:b_motion_axial} with the equilibrium DW width $\Delta=\Delta^b_0$.

\section{Magnetic nanotube with DMI of interfacial type}\label{app:iDMI}

\subsection{Static DW solution}\label{app:iDW_sol}

In the no-driving case $h=0$, the ground state is defined by the set of static equations $\delta\mathcal{E}^i/\delta\Theta=0$ and $\delta\mathcal{E}^i/\delta\Phi=0$, which read:
\begin{widetext}
\begin{equation}\label{eq:iDMI_euler}
	\begin{split}		&-2\left(\partial_{11}\Theta+\partial_{22}\Theta\right)+\sin2\Theta\left[\left(\partial_1\Phi\right)^2+\left(\partial_2\Phi\right)^2+\mathcal{K}_1^i+\mathcal{K}_3^i\sin^2\Phi\right]-2\mathcal{D}_{13}^{i(1)}\cos\Phi\sin^2\Theta\partial_1\Phi+\mathcal{D}_{23}^{i(2)}\sin2\Theta\partial_2\Phi=0,\\		&-2\left[\partial_1\left(\sin^2\Theta\partial_1\Phi\right)+\partial_2\left(\sin^2\Theta\partial_2\Phi\right)\right]+\mathcal{K}_3^i\sin2\Phi\sin^2\Theta+2\mathcal{D}_{13}^{i(1)}\cos\Phi\sin^2\Theta\partial_1\Theta - \mathcal{D}_{23}^{i(2)}\sin2\Theta\partial_2\Theta=0.
	\end{split}
\end{equation}
\end{widetext}
Similarly to Sec.~\ref{app:bDW_sol}, for the case of DMI-free nanotube the ground state of the system is doubly degenerated: $\Theta=0$ and $\Theta=\pi$. The DW structure is defined by~\eqref{eq:dw_dmi_free}:
\begin{equation}
	\Theta_\textsc{dw}^0\left(\xi_2\right) = 2\arctan e^{p \xi_2/\Delta}, \quad \cos\Phi_\textsc{dw}^0 = \pm 1,
\end{equation}

The DW solution~\eqref{eq:dw_dmi_free} does not satisfy Eqs.~\eqref{eq:iDMI_euler} in case $d\neq 0$. Here, we will also introduce small deviations~\eqref{eq:dw_deformed} from the non-perturbed solution. Substituting now~\eqref{eq:dw_deformed} into~\eqref{eq:iDMI_euler} and linearizing the obtained equation with respect to the deviations, we obtain the following
equations for the deviations $\vartheta$ and $\varphi$:
\begin{subequations}\label{eq:dw_i_deformed}
	\begin{align} \label{eq:dw_i_deformed-vartheta}
		&\vartheta'' + \left(\frac{2}{\cosh^2\zeta}-1\right)\vartheta=0,\\
		\label{eq:dw_i_deformed-varphi}
		&\varphi'' +\left(\frac{2}{\cosh^2\zeta}-\alpha^2\right)\varphi = \beta^i\frac{\tanh\zeta}{\cosh\zeta},
	\end{align}
\end{subequations}
where $\beta^i =  p\,\mathcal{D}_{23}^{i(2)}/{\sqrt{\mathcal{K}_1^b}}$. Equation~\eqref{eq:dw_i_deformed-vartheta} coincides with Eq.~\eqref{eq:dw_b_deformed-vartheta}, while the function $\varphi(\zeta)$ is also exponentially localized, $\varphi \propto \beta^i \exp(-\alpha\zeta)$ when $\zeta\to\infty$ but has linear behavior $\varphi\propto\beta^i\zeta/(2-\alpha^2)$ when $\zeta\to0$.

The set of equations~\eqref{eq:dw_i_deformed} was previously obtained for DWs in straight biaxial wires with bulk-type DMI~\cite{Kravchuk14}. 

\subsection{Magnetostatic energy}\label{app:iDW_ms}

For the case of a DW in a nanotube with interfacial DMI given by ansatz~\eqref{eq:i_dw_solution}, the surface-charge-induced magnetostatic energy term~\eqref{eq:MSthin}, takes the form:
\begin{equation}
E_{\mathrm{MS}}^{\sigma -\sigma } \simeq \frac{1}{2}\mu _{0}M_{s}^{2} 2\pi R W \Delta \left(1-\pi a\frac{\cos2\varPhi}{\sinh\left(\pi a\right)}\right), 
\end{equation}
which can be accounted for by the renormalization of the anisotropy constant $\mathcal{K}^i_3$ in~\eqref{eq:i_const} through shifting the anisotropy ratio as $\varepsilon=\epsilon+Q^{-1}$.   

The volume and geometric magnetostatic charges for the magnetization distribution given by~\eqref{eq:i_dw_solution} are: 
\begin{eqnarray}
\rho ^{i} &=&\frac{1}{\Delta }\frac{
a\sin \left(
\varPhi _{0}+a\frac{\xi _{2}}{\Delta} \right) +\tanh \frac{\xi _{2}}{\Delta} \cos \left(
\varPhi _{0}+a\frac{\xi _{2}}{\Delta} \right) }{\cosh \frac{\xi _{2}}{\Delta}};  \nonumber \\
g^{i} &=&-\frac{1}{\xi _{3}}\frac{\sin \left(
\varPhi _{0}+a\frac{\xi _{2}}{\Delta} \right)}{\cosh \frac{\xi _{2}}{\Delta}}. 
\end{eqnarray}

Note that in the case of interfacial-type DMI, the magnetization components $m_2$ and $m_3$ have different parity with respect to $\xi_2\rightarrow -\xi_2$. This leads to the emergence of the DW configuration with a favorable phase slope $a$, as both contributions ${E}_\textsc{ms}^{g-\rho}$ and ${E}_\textsc{ms}^{\sigma-\rho}$ embrace the term $\propto a \cos\varPhi _{0}^2$. However, the DW topological charge $p=\pm 1$ defining the component $m_1$~\eqref{eq:i_dw_solution} plays no role in magnetostatic energy of the studied system, which means the chiral effects are absent. The preferable phase slope $a$ is the same for any $p$ and any DW helicity $\mathcal{C} = \cos\varPhi _{0} =\pm 1$. 

In the local interactions framework, the preferable sign of phase slope $a$ is pre-determined by the intrinsic DMI constant $d$ of the material, $\sign a = \sign d$. However, the magnetostatic contribution prefers a particular sign of $a$ independent of $d$.
The resulting configuration depends on the interplay of geometrical and material parameters in a specific system, as we proceed to derive. 

The total magnetostatic energy can be written as:
\begin{widetext}
\begin{equation}
E_{\mathrm{MS}}=\frac{1}{2}\mu _{0}M_\text{s}\pi \int_{0}^{\infty }\left\{
k^{2}G_{1}^{2}\left( k\right) \Lambda _{1}-kG_{1}\left( k\right) G_{0}\left(
k\right) \left[ \Lambda _{2}-\widetilde{\Lambda }_{1}\right]
-G_{0}^{2}\left( k\right) \widetilde{\Lambda }_{2}\right\} \textrm{d}k,
\end{equation}
\end{widetext}
where we use the notations a l{\`a}~\cite{Landeros10}:

$G_{\nu }\left( k\right) =\int\limits_{R_{1}}^{R_{2}}J_{\nu }\left( kx_{3}\right)
x_{3}\textrm{d}x_{3}$;

$\Lambda _{1}=\int\limits_{-\infty }^{\infty }\textrm{d}x_{2}\,m_{3}\left( x_{2}\right)
\int\limits_{-\infty }^{\infty }\textrm{d}x_{2}^{\prime }e^{-k\left\vert x_{2}-x_{2}^{\prime
}\right\vert }m_{3}\left( x_{2}^{\prime }\right) $; \ 

$\widetilde{\Lambda }_{1}=\int\limits_{-\infty }^{\infty }\textrm{d}x_{2}\,m_{3}\left(
x_{2}\right) \int\limits_{-\infty }^{\infty }\textrm{d}x_{2}^{\prime }e^{-k\left\vert
x_{2}-x_{2}^{\prime }\right\vert }\partial _{x_{2}^{\prime }}m_{2}\left(
x_{2}^{\prime }\right) $;

$\Lambda _{2}=\int\limits_{-\infty }^{\infty }\textrm{d}x_{2}\,m_{2}\left( x_{2}\right)
\partial _{x_{2}} \int\limits_{-\infty }^{\infty }\textrm{d}x_{2}^{\prime
}e^{-k\left\vert x_{2}-x_{2}^{\prime }\right\vert }m_{3}\left( x_{2}^{\prime
}\right)  $; 

$\widetilde{\Lambda }_{2}=\int\limits_{-\infty }^{\infty }\textrm{d}x_{2}\,m_{2}\left(
x_{2}\right) \partial _{x_{2}} \int\limits_{-\infty }^{\infty }\textrm{d}x_{2}^{\prime
}e^{-k\left\vert x_{2}-x_{2}^{\prime }\right\vert }\partial _{x_{2}^{\prime
}}m_{2}\left( x_{2}^{\prime }\right)  $.

The energy contributions ${E}_\textsc{ms}^{\sigma-\rho}$ and ${E}_\textsc{ms}^{g-\rho}$ are captured by the terms containing $\widetilde{\Lambda }_{1}$ and $\Lambda _{2}$. Using~\eqref{eq:i_dw_solution}, applying  $e^{-k\left\vert x_{2}-x_{2}^{\prime }\right\vert }=\frac{1}{2\pi }%
\int_{-\infty }^{\infty }\frac{2k}{k^{2}+q^{2}}e^{iq\left(
x_{2}-x_{2}^{\prime }\right) }\textrm{d}q$, and leaving only the terms linear in $a$, we get:
\begin{equation}
{E}_\textsc{ms}^{\sigma-\rho}+{E}_\textsc{ms}^{g-\rho} \simeq -\frac{\mu _{0}M_\text{s}}{4}a\pi^2 W^{2}\Delta.
\label{eq:MSinterf}
\end{equation}

Therefore, for any DW helicity the configuration with $a>0$ is preferred by magnetostatics. We see from~\eqref{eq:i_dw_energy} and~\eqref{eq:MSinterf} that for $d>0$ both DMI and magnetostatics favour the configuration with $a>0$ (corresponds to Fig.~\ref{fig:iDW}(c),(d)). However, if $d<0$ (Fig.~\ref{fig:iDW}(e)), the two effects compete. The effective DMI constant $d+{\pi w \varkappa}/{8Q}$ defines the resulting phase slope: $d<-{\pi w \varkappa}/{8Q}$ results in $a<0$, while $d>-{\pi w \varkappa}/{8Q}$ would still favor $a>0$. Interestingly, the threshold DMI value depends on the film thickness $w$ and curvature $\varkappa$, which enables tailoring the interplay between local and nonlocal effects on the DW profile at the nanotube production stage. In a nanotube with $d<0$ and $W/R=-8Qd/\pi$ the DMI-induced effects can be virtually compensated, and the resulting DW structure will be described by~\eqref{eq:dw_dmi_free}.

\subsection{Exact form of the equations of motion for DW}\label{app:iDW_exat}

By substituting the Ansatz~\eqref{eq:i_q-phi} into~\eqref{eq:LR_functions} one obtains
\begin{eqnarray}\label{eq:i_LR}
		\frac{\mathcal{L}_\textsc{dw}^i}{2\pi/\varkappa}& = &2p\left[\varPhi\dot{q}+ca\dot{\Delta}\right]  - \frac{\mathcal{E}_\textsc{dw}^i}{2\pi/\varkappa},\\
		\frac{\mathcal{R}_\textsc{dw}^i}{2\pi/\varkappa}& =&\frac{\eta}{\Delta}\left[\dot{q}^2+\left(\dot{q}a-\dot{\varPhi}\Delta\right)^2+c\left(\Delta \dot{a}-\dot{\Delta}a\right)^2+c\dot{\Delta}^2\right], \nonumber 
\end{eqnarray}
with total energy in an external magnetic field
\begin{equation}\label{eq:i_energyZ}
	\begin{split}
		\frac{\mathcal{E}_\textsc{dw}^i}{2\pi/\varkappa}& = \frac{2}{\Delta}\left(1+a^2\right)+2a\mathcal{D}_{23}^{i(2)}-a \Delta \frac{\pi w \varkappa }{4Q}\\
		 +\Delta&\left[2\mathcal{K}_1^i+\mathcal{K}^i_3\left(1-\pi a\frac{\cos2\varPhi}{\sinh\left(\pi a\right)}\right)\right]-4phq.
	\end{split}
\end{equation}

Lagrangian and dissipative function~\eqref{eq:i_LR} produce the following set of equations of motion:
\begin{widetext}
\begin{equation}\label{eq:iDW_equation_of_motion}
		\begin{split}
			&\eta\left(1+a^2\right)\frac{\dot{q}}{\Delta}+\left(p-\eta a\right)\dot{\varPhi} = 2ph,\\
			&\left(p+\eta a\right)\dot{q}-\eta\Delta\dot{\varPhi} = \pi a\Delta\mathcal{K}^i_3\frac{\sin2\varPhi}{\sinh\left(\pi a\right)},\\
			&c\eta\left(1+a^2\right)\frac{\dot{\Delta}}{\Delta}+c\left(p-\eta a\right)\dot{a} = \frac{1}{\Delta^2}\left(1+a^2\right)-\mathcal{K}_1^i-\frac{\mathcal{K}^i_3}{2}\left[1-\pi a\frac{\cos2\varPhi}{\sinh\left(\pi a\right)}\right] + a\frac{\pi w \varkappa}{8Q},\\
			&c\left(p+\eta a\right)\frac{\dot{\Delta}}{\Delta}-c\eta\dot{a} =\frac{2}{\Delta}\left(\frac{a}{\Delta}+\frac{\mathcal{D}_{23}^{i(2)}}{2}-\frac{\pi w \varkappa \Delta}{16Q}\right)+\mathcal{K}^i_3\frac{\pi}{2}\frac{\cos2\varPhi}{\sinh\left(\pi a \right)}\left[\pi a\coth\left(\pi a\right)-1\right].
		\end{split}
\end{equation}
\end{widetext}
In linear approximation with respect to the slope $a$,
\begin{equation}\label{eq:iDW_equation_of_motion_lin}
	\begin{split}
		&\eta\frac{\dot{q}}{\Delta}+ p\left(1-a\eta\right)\dot{\varPhi} = 2ph,\\
		&p\left(1+a\eta\right)\dot{q}-\eta\Delta\dot{\varPhi} = \mathcal{K}_{3}^i\Delta\sin2\varPhi,\\
		&c\eta \frac{\dot{\Delta}}{\Delta}+ cp\dot{a}=\frac{1}{\Delta^2}-\mathcal{K}_1^i-\mathcal{K}^i_3\sin^2\varPhi + a\frac{\pi w \varkappa}{8Q}, \\
		&\left(p+\eta a\right)\dot{\Delta}-\eta\dot{a}\Delta=\frac{1}{c}\left(\mathcal{D}_{23}^{i(2)}-\Delta \frac{\pi w \varkappa}{8Q}+\frac{2a}{\Delta}\right)\\
        &+2a\mathcal{K}^i_3\Delta\cos2\varPhi.
	\end{split}
\end{equation}
The third and fourth equations in~\eqref{eq:iDW_equation_of_motion_lin} demonstrate that $\Delta$ and $a$ relax towards their equilibrium values $\Delta_0^i=1/\sqrt{\mathcal{K}_1^i+\mathcal{K}^i_3\sin^2\varPhi^i}$ and $a_0^i = -\left(\mathcal{D}_{23}^{i(2)}-\frac{\pi w \varkappa }{8Q}\Delta_0^i\right)/\left(2/\Delta_0^i+2c\mathcal{K}^i_3\Delta_0^i\cos2\varPhi^i\right)$. The characteristic time of this relaxation is proportional to the damping $\propto \eta$~\cite{Hillebrands06}. We conclude that the DW width and phase slope are slave variables, $\Delta\left[\varPhi(t)\right]$ and $a\left[\varPhi(t)\right]$, and the DW dynamics can be described by~\eqref{eq:i_motion_vortex} with the equilibrium values of $\Delta=\Delta^i_0$ and $a=a_0^i$.

\section{Detail of numerical simulations}\label{app:sim}

In order to verify our analytical calculations we perform a set numerical simulations with (i) OOMMF code~\cite{OOMMF} with additional packages for bulk-type DMI~\cite{Cortes-Ortuno18,Cortes-Ortuno18a} and a modified package for interfacial-type DMI, and (ii) in-house developed python-based code. The modified package for OOMMF is based on the module developed by S.~Rohart \textit{et al.}~\cite{Rohart13}. The modified package is in Supplemental Materials~\cite{Note2}.

\subsection{OOMMF simulations}
With OOMMF we consider only static problems~\cite{Note1}. In simulations we consider tubes of total length 250 nm, inner radius 24 nm,  and thickness 3 nm, with cell size $1\times 1\times 1$ nm$^3$. We use the following artificial material parameters: exchange constant $A = 25\times 10^{-12}$ J/m, saturation magnetization $M_s = 5 \times 10^5$ A/m, easy-axial anisotropy constant $K_a = 10^6$ J/m$^3$, easy-surface anisotropy constant $K_s = 0.34\times10^6$ J/m$^3$, and DMI constant $D=1.25\times 10^{-3}$ J/m$^2$. The effective easy-surface anisotropy constant with the account of magnetostatics was set to $K_s^\text{eff} = K_s + \mu_0 M_s^2 /2 = 0.5\times10^6$ J/m$^3$.

The numerical experiment consists of two steps. First, we relax DW structure in an overdamped regime ($\eta = 0.5$) with a certain topological charge $p=+1$, i.e. we start with two sharp domain which are oriented in opposite directions. Next, we extract the DW parameters: DW position, phase, width, and phase slope. The results of numerical simulations are presented in Figs.~\ref{fig:bDW}--\ref{fig:vel_iDMI}.

\subsection{Spin-lattice simulations}
We consider a cylindrical surface as a square lattice with lattice constant $s$. Each node is characterized by a magnetic moment $\vec{m}_{\vec{k}}(t)$ which is located at the position $\vec{r}_{\vec{k}}$. Here $\vec{k} = (i, j)$ is a two dimensional vector that defines the magnetic moment and its position on the lattice with size $N_1\times N_2$ ($i\in[1, N_1]$ and $j \in [1, N_2]$). Magnetic moments are ferromagnetically coupled. We are interested in the case when the system is a closed cylindrical surface, hence we impose the periodical boundary conditions $\vec{m}_{(N_1+1, j)} = \vec{m}_{(1, j)}$ and $\vec{r}_{(N_1+1, j)} = \vec{r}_{(1, j)}$. The dynamics of magnetic system is govern by discrete Landau--Lifshitz--Gilbert equations. 

The dynamical problem is considered as a set of $3N_1N_2$ ordinary differential equations with respect to $3N_1N_2$ unknown functions $m^\textsc{x}_{\vec{k}}(t),\ m^\textsc{y}_{\vec{k}}(t),\ m^\textsc{z}_{\vec{k}}(t)$. For given initial conditions, the set of Landau--Lifshitz--Gilbert equations is integrated numerically using Runge--Kutta method in Python.  For details of workflow of in-house developed code see Ref.~\cite{Yershov20}.

In simulations we consider cylinders with $N_1 = 200$ and $N_2 = 10000$, the magnetic length $\ell \approx 6.37s$ (this value of magnetic length results in dimensionless curvature $\varkappa = 0.2$), DMI constant $d = 0.25$ (for bulk and interfacial DMI types), and anisotropy ratio $\varepsilon = 0.5$. 

The numerical experiment consists of two steps. First, we relax the DW structure in an over-damped regime $(\eta = 0.5)$ to relax the DW of the certain charge $p$ and helicity $\mathcal{C}$ on the cylindrical surface. In the second step, we stimulate the DW motion by applying the magnetic field $h$ with natural damping coefficient $\eta = 0.01$. The averaged DW velocity is obtained as
$V = \frac{1}{T}\int_0^{T}\dot{q}\mathrm{d}t$,
where $\dot{q}$ is extracted from the simulation data. 

\subsection{Movies of the domain wall motion}
For better visualization of the field-induced motion of magnetic DWs in a nanotube with bulk-type DMI, we have prepared movies~\cite{Note2} illustrating the wall dynamics:
\begin{itemize}
	\item \texttt{video1\_favorable\_DW.mp4} shows the dynamics of a DW with topological charge $p=+1$ and initial helicity $\mathcal{C}=+1$;
	\item \texttt{video2\_unfavorable\_DW.mp4} shows the dynamics of a DW with topological charge $p=+1$ and initial helicity $\mathcal{C}=-1$.
\end{itemize}

The data in these movies correspond to results obtained via numerical simulations. In all simulations, we used $d=0.25$, $\varkappa=0.25$, $h=0.7h_\textsc{w}^0$, and $\eta = 0.01$. 

In both videos, the bottom right panel displays a magnified view of the nanotube near the DW center, located between two green circles. The top left panel presents an unrolled perspective of this zoomed-in region where the color scheme corresponds to magnetization component $\mu_2$, and streamlines show the in-surface magnetization.

%\bibliography{papers_lib}

\end{document}